%% file: ms.tex
\documentclass{article}

\input{ms_preamble}

\begin{document}

\subfile{ms_introduction}

\subfile{ms_adaptingkernels}

\subfile{ms_directfptapproach}

\subfile{ms_oct}

\subfile{ms_lowerbounds}

\subfile{ms_conclusion}

\bibliographystyle{plain}
\bibliography{bib}

\end{document}

%% file: ms_preamble.tex
\usepackage{subfiles,graphicx,amsmath,amssymb,amsthm,color,thmtools,hyperref,environ,mathtools,enumitem}
\usepackage[paper=a4paper,margin=1in]{geometry}
\usepackage{authblk}
        
\usepackage{algorithm}
\usepackage[noend]{algpseudocode}
\algnewcommand\algorithmicforeach{\textbf{for each}}
\algdef{S}[FOR]{ForEach}[1]{\algorithmicforeach\ #1\ \algorithmicdo}

\declaretheorem[style=definition,numberwithin=section]{definition}

\declaretheorem[style=plain,sibling=definition]{theorem}
\declaretheorem[style=plain,sibling=definition]{lemma}
\declaretheorem[style=plain,sibling=definition]{proposition}
\declaretheorem[style=plain,sibling=definition]{corollary}

\newcommand{\N}{\ensuremath{\mathbb{N}}}
\renewcommand{\O}{\ensuremath{\mathcal{O}}}
\newcommand{\Ot}{\ensuremath{\tilde{\mathcal{O}}}}

\newcommand{\Ftwo}{\ensuremath{\mathbb{F}_2}}
\newcommand{\M}{\ensuremath{\mathcal{M}}}

\newcommand{\U}{\ensuremath{\mathcal{U}}}
\newcommand{\X}{\ensuremath{\mathcal{X}}}
\newcommand{\Y}{\ensuremath{\mathcal{Y}}}
\newcommand{\A}{\ensuremath{\mathcal{A}}}

\renewcommand{\qed}{\ensuremath{\hfill\square}}

\makeatletter
\newcommand{\problemtitle}[1]{\gdef\@problemtitle{#1}}
\newcommand{\probleminput}[1]{\gdef\@probleminput{#1}}
\newcommand{\problemquestion}[1]{\gdef\@problemquestion{#1}}
\newcommand{\problemparameter}[1]{\gdef\@problemparameter{#1}}
\NewEnviron{problemP}{
	\problemtitle{}\probleminput{}\problemquestion{}
	\BODY
	\par\addvspace{.5\baselineskip}
	\noindent
	\centering{
	\fbox{\parbox{.9\textwidth}{
		\@problemtitle \\
		\textit{Input: } \@probleminput \\
		\textit{Question: }	\@problemquestion
	}}}
	\par\addvspace{.5\baselineskip}
}

\NewEnviron{problemParam}{
	\problemtitle{}\probleminput{}\problemquestion{}\problemparameter{}
	\BODY
	\par\addvspace{.5\baselineskip}
	\noindent
	\centering{
	\fbox{\parbox{.9\textwidth}{
			\@problemtitle \hfill\\%
			\textit{Input: } \@probleminput \\
			\textit{Parameter: } \@problemparameter \\
			\textit{Question: }	\@problemquestion
	}}}
	\par\addvspace{.5\baselineskip}
}
\makeatother

\newcommand{\problemISI}{\textsc{Induced Subgraph Isomorphism}}
\newcommand{\problemPC}{\textsc{$k \times k$ Permutation Clique}}
\newcommand{\problemPfDV}{\textsc{$\Pi$-free Deletion~[VC]}}

%% file: ms_introduction.tex
\title{Streaming Deletion Problems Parameterized by Vertex Cover\thanks{This work is based on the master thesis "Parameterized Algorithms in a Streaming Setting" by the first author. This work is partially supported by the NWO grant OCENW.KLEIN.114 (PACAN). The authenticated conference publication is available online at \url{https://doi.org/10.1007/978-3-030-86593-1_29}.}}

\author[1]{Jelle J. Oostveen\thanks{j.j.oostveen@uu.nl}}
\author[1]{Erik Jan van Leeuwen\thanks{e.j.vanleeuwen@uu.nl}}
\affil[1]{Dept.\ Information and Computing Sciences, Utrecht University, The Netherlands.}

\maketitle
\begin{abstract}
Streaming is a model where an input graph is provided one edge at a time, instead of being able to inspect it at will. In this work, we take a parameterized approach by assuming a vertex cover of the graph is given, building on work of Bishnu et al.\ [COCOON~2020]. We show the further potency of combining this parameter with the Adjacency List streaming model to obtain results for vertex deletion problems. This includes kernels, parameterized algorithms, and lower bounds for the problems of \textsc{$\Pi$-free Deletion}, \textsc{$H$-free Deletion}, and the more specific forms of \textsc{Cluster Vertex Deletion} and \textsc{Odd Cycle Transversal}. We focus on the complexity in terms of the number of passes over the input stream, and the memory used. This leads to a pass/memory trade-off, where a different algorithm might be favourable depending on the context and instance. We also discuss implications for parameterized complexity in the non-streaming setting.
\end{abstract}

\section{Introduction}
Streaming is an algorithmic paradigm to deal with data sets that are too large to fit into main memory \cite{HenzingerStreams}. Instead, elements of the data set are inspected in a fixed order\footnote{We consider insertion-only streams throughout this paper.} and aggregate data is maintained in a small amount of memory (much smaller than the total size of the data set). It is possible to make multiple passes over the data set. The goal is to design algorithms that analyze the data set while minimizing the combination of the number of passes and the required memory. We note that computation time is not measured in this paradigm. Streaming has proved very successful and is extensively studied in many diverse contexts \cite{StreamingModels2_Survey,MuthukrishnanStreams}. In this work, we focus on the case where the data sets are graphs and the streamed elements are the edges of the graph.

A significant body of work on graph streaming works in the semi-streaming model, where $\Ot(n)$ memory\footnote{Throughout this paper, memory is measured in bits. The $\Ot$ notation hides factors polylogarithmic in $n$. Note that $\O(\log n)$ bits is the space required to store (the identifier of) a single vertex or edge.} is allowed, with the aim of limiting the number of necessary passes to one or two. This memory requirement might still be too much for the largest of networks. Unfortunately, many basic problems in graphs require $\Omega(n)$ or even worse space~\cite{OmegaNMemory1,OmegaNMemory2} to compute in a constant number of passes. Therefore, Fafianie and Kratsch~\cite{kratsch} and Chitnis et al.~\cite{ChitnisMaxMatchVC} introduced concepts and analysis from parameterized complexity \cite{DowneyFellowsBook} to the streaming paradigm. For example, it can be decided whether a graph has a vertex cover of size at most $K$ using one pass and $\Ot(K^2)$ space, which is optimal. This led to various further works \cite{ChitnisEsfandiariSampling,SaketVCNrConferenceVersion,MinOnesdSAT} and the first systematic study by Chitnis and Cormode \cite{ChitnisTheory}.

Our work continues this line of research and follows up on recent work by Bishnu et al. \cite{SaketVCNrConferenceVersion,SaketVCNr}\footnote{As the Arxiv version contains more results, we refer to this version from here on.}. They made two important conceptual contributions. First, they analyzed the complexity of parameterized streaming algorithms in three models that prescribe the order in which the edges arrive in the stream and that are commonly studied in the literature~\cite{SaketVCNrConferenceVersion,StreamingModels1,StreamingModels2_Survey,StreamingModels3}. The \emph{Edge Arrival} (EA) model prescribes some permutation of all the edges of the graph. The \emph{Vertex Arrival} (VA) requires that the edges appear per vertex: if we have seen the vertices $V' \subseteq V$ already, and the next vertex is $w$, then the stream contains the edges between $w$ and the vertices in $V'$. Finally, the \emph{Adjacency List} (AL) gives the most information, as it requires the edges to arrive per vertex, but when vertex $v$ appears in the stream, we also see all edges incident to $v$. This means we effectively see every edge twice in a single pass, once for both of its endpoints.

The second and more important contribution of Bishnu et al.~\cite{SaketVCNr} was to study the size $K$ of a vertex cover in the graph as a parameter. This has been broadly studied in parameterized complexity (see e.g.\ the PhD thesis of Jansen~\cite{jansenBook}). They showed that the very general \textsc{$\mathcal{F}$-Subgraph Deletion} and \textsc{$\mathcal{F}$-Minor Deletion} problems all admit one pass, $\Ot(\Delta(\mathcal{F})\cdot K^{\Delta(\mathcal{F})+1})$ space streaming algorithms in the AL model, by computing small kernels to which then a straightforward exhaustive algorithm is applied. On the other hand, such generic streaming algorithms are not possible in the EA and VA models, as then (super-) linear lower bounds exist even if the size of a smallest vertex cover is constant~\cite{SaketVCNr}.

We focus on the induced subgraph version of the vertex deletion problem, parameterized by the size of a vertex cover. Here, $\Pi$ is a collection of graphs.

\begin{problemParam}
	\problemtitle{\problemPfDV{}}
	\probleminput{A graph $G$ with a vertex cover $X$, and an integer $\ell \geq 1$.}
	\problemparameter{The size $K \coloneqq |X|$ of a vertex cover.}
	\problemquestion{Is there a set $S\subseteq V(G)$ of size at most $\ell$ such that $G[V(G)\setminus S]$ does not contain a graph in $\Pi$ as an induced subgraph?}
\end{problemParam}
To avoid triviality\footnote{Otherwise, removing the entire vertex cover is a trivial solution.}, we assume every graph in $\Pi$ is edgeless or $K \geq \ell$. We assume the vertex cover is given as input; if only the size is given, we can use one pass and $\Ot(K^2)$ space or $2^K$ passes and $\Ot(K)$ space to obtain it \cite{ChitnisMaxMatchVC,ChitnisTheory} (this does not meaningfully impact our results).
The unparameterized version of this problem is well known to be NP-hard~\cite{PiDeletionNP} for any nontrivial and hereditary property $\Pi$. It has also been well studied in the parameterized setting (see e.g.~\cite{CaiModification,PiDeletionFOlogic,SauS20}). When parameterized by the vertex cover number, it has been studied from the perspective of kernelization: while a polynomial kernel cannot exist in general~\cite{BodlaenderJK14,jansenPaper}, polynomial kernels exist for broad classes of families~$\Pi$~\cite{jansenPaper,JansenKroonNewPiFreeDeletion}. As far as we are aware, parameterized algorithms for this parameterization have not been explicitly studied.

In the streaming setting, Chitnis et al. \cite{ChitnisEsfandiariSampling} showed for the unparameterized version of this problem in the EA model that any $p$-pass algorithm needs $\Omega(n/p)$ space if $\Pi$ satisfies a mild technical constraint. 
For some \problemPfDV{} problems, the results by Bishnu et al.~\cite{SaketVCNr} imply single-pass, $\mbox{poly}(K)$ space streaming algorithms (through their kernel for \textsc{$\mathcal{F}$-Subgraph Deletion} [VC]) in the AL model and lower bounds in the EA/VA model. They also provide an explicit kernel for \textsc{Cluster Vertex Deletion [VC]} in the AL/EA/VA models. However, this still leaves the streaming complexity of many cases of the \problemPfDV{} problem open.

\paragraph{Our contributions} 
We determine the streaming complexity of the general \problemPfDV{} problem. Our main positive result is a unified approach to a single-pass polynomial kernel for \problemPfDV{} for a broad class of families $\Pi$. 
In particular, we show that the kernelization algorithms by Fomin et al.~\cite{jansenPaper} and Jansen and Kroon~\cite{JansenKroonNewPiFreeDeletion} can be adapted to the streaming setting. The kernels of Fomin et al. \cite{jansenPaper} consider the case when $\Pi$ can be characterized by few adjacencies, which intuitively means that for any vertex of any member of $\Pi$, adding or deleting edges between all but a few (say at most $c_\Pi$) distinguished other vertices does not change membership of $\Pi$. The exponent of the polynomial kernels depends on $c_\Pi$. Jansen and Kroon~\cite{JansenKroonNewPiFreeDeletion} considered even more general families $\Pi$. We show that these kernels can be computed in the AL model using a single pass and polynomial space (where the exponent depends on $c_\Pi$). This generalizes the previous results by Bishnu et al.~\cite{SaketVCNr} as well as their kernel for \textsc{$\mathcal{F}$-Subgraph Deletion} [VC].

To complement the kernels, we take a direct approach to find more memory-efficient algorithms, at the cost of using many passes.
We show novel parameterized streaming algorithms that require $\Ot(K^2)$ space and $\O(K)^{\O(K)}$ passes. Here, all hidden constants depend on $\Pi$ and $c_\Pi$. Crucially, however, the exponent of the space usage of these algorithms does not, which provides an advantage over computing the kernel. We also provide explicit streaming algorithms for \textsc{Cluster Vertex Deletion}~[VC] and \textsc{Odd Cycle Transversal}~[VC] that require $\Ot(K)$ space (both) and $2^K K^2$ and $3^K$ passes respectively, as well as streaming algorithms for \textsc{$\Pi$-free Deletion}~[VC,$|V(H)|$] when $\Pi = \{H\}$ and the problem is parameterized by $K$ and $|V(H)|$. A crucial ingredient to these algorithms is a streaming algorithm that finds induced subgraphs isomorphic to a given graph $H$. For these algorithms, a $\Pi$ of large size can lead to large memory use. To this end, we also give algorithms using an oracle to learn information about $\Pi$, similar to an algorithm by Cai~\cite{CaiModification}. Further details are provided in Section~\ref{sec:DirectFPT}.

The above results provide a trade-off in the number of passes and memory complexity of the algorithm used. However, we should justify using both the AL model and the parameter vertex cover. To this end, in Section~\ref{sec:lowerbounds}, we investigate lower bounds for streaming algorithms for \problemPfDV{}. The (unparameterized) linear lower bound of Chitnis et al. \cite{ChitnisEsfandiariSampling} in the EA model requires that $\Pi$ contains a graph $H$ for which $|E(H)| \geq 2$ and no proper subgraph is a member of $\Pi$. We prove that the lower bound extends to both the VA and AL models, with only small adjustments. Hence, parameterization is necessary to obtain sublinear passes and memory for most $\Pi$. Since \textsc{Vertex Cover} is one of the few natural graph parameters that has efficient parameterized streaming algorithms~\cite{kratsch,ChitnisMaxMatchVC}, this justifies the use of the vertex cover parameter. We also extend the reductions by Bishnu et al.~\cite{SaketVCNr} to general hardness results for \textsc{$\Pi$-free Deletion} in the VA and EA model when the size of the vertex cover is a constant (dependent on~$\Pi$), justifying the use of the AL model for most $\Pi$.

We also consider the parameterized complexity of \textsc{$H$-free Deletion}~[VC] in the non-streaming setting. While polynomial kernels were known in the non-streaming setting \cite{jansenPaper}, we are unaware of any investigation into explicit parameterized algorithms for these problems. We give a general $2^{O(K^2)} \mathrm{poly}(n,|V(H)|)$ time algorithm. This contrasts the situation for \textsc{$H$-free Deletion} parameterized by the treewidth~$t$ of the graph, where a $2^{o(t^{|V(H)|-2})} \mathrm{poly}(n,|V(H)|)$ time lower bound is known under the Exponential Time Hypothesis (ETH)~\cite{SauS20}. We also construct a graph property $\Pi$ for which we provide a lower bound of $2^{o(K \log K)} \mathrm{poly}(n,|V(H)|)$ for \problemPfDV{} under ETH. Further details are provided in Section~\ref{sec:DirectFPT}.

\paragraph{Preliminaries.} We work on undirected graphs $G = (V,E)$, where $|V| = n, |E| = m$. We denote an edge $e \in E$ between $v \in V$ and $u \in V$ with $uv \in E$. For a set of vertices $V' \subseteq V$, denote the subgraph induced by $V'$ as $G[V']$. Denote the neighbourhood of a vertex $v$ with $N(v)$ and for a set $S$ denote $N(S)$ as $\bigcup_{v \in S}N(v)$. We write $N[v]$ for $N(v)$ including $v$, so $N[v] = N(v) \cup \{v\}$.

We denote the parameters of a problem in $[\cdot]$ brackets, a problem $\textsc{A}$ parameterized by vertex cover number and solution size is denoted by \textsc{A~[VC, $\ell$]}.

\paragraph{Further Related Work.} Our main algorithm for \problemPfDV{} uses a procedure that finds an occurrence of an induced subgraph $H$ in the input graph $G$. To this end, there is related work on the \textsc{Induced Subgraph Isomorphism} problem. In general, finding an induced copy of a graph $H$ of size $k$ can be brute-forced in $\O(n^k)$ time. Ne{\v{s}}et{\v{r}}il and Poljak~\cite{nevsetvril1985complexity} show that finding an induced copy of a graph $H$ can be reduced in time $\O(k^2n^2)$ to the \textsc{$k$-Clique} or \textsc{$k$-Independent Set} problem, which makes a $k$-clique and a $k$-independent set essentially the hardest patterns to detect. With this in mind, it is logical that knowing a vertex cover of the input graph can help in our approach. Dalirrooyfard et al.~\cite{DalirrooyfardVW19} relate the hardness of \textsc{Induced Subgraph Isomorphism} to the hardness of \textsc{$k$-Clique}, which holds under the Hadwiger conjecture. Under ETH, this implies that for almost all $H$ of size $k$, finding an $H$ in a graph $G$ cannot be done in time $n^{o(k/\log k)}$.

In terms of upper bounds, the running time of detecting a $k$-clique is related to fast matrix multiplication~\cite{ItaiR78,nevsetvril1985complexity,EisenbrandG04}. There is also work on \textsc{Induced Subgraph Isomorphism} on specific graph classes \cite{konagaya2016polynomial,kijima2012subgraph}, and on specific small patterns $H$ \cite{CorneilPS85,Olariu90a,EisenbrandG04,KloksKM00,KowalukLL13,RegularSubgraphFind,FindCountSmallSubgraphsEfficiently,WilliamsWWY15,BlaserKS18}. Eppstein~\cite{eppstein2002subgraph} investigates the parameterized complexity of the \textsc{Induced Subgraph Isomorphism} problem with the parameter treewidth, and shows an algorithm listing all occurrences of $H$ of size $w$ that contain a vertex of a set $S \subseteq V(G)$ in time $2^{\O(w \log w)}n + \O(kw)$ for graphs of treewidth $w$, where $k$ is the number of isomorphisms. This result is related to our algorithm in that a vertex cover of size $k$ implies bounded treewidth, and our running time is similar. However, our result is more general because we allow freedom in the size of $H$ as compared to the graph parameter.

\subsection{Memory complexity of branching algorithms}\label{sec:memcomplbranching}

In this work, we will see branching algorithms which branch on, for example, including a certain vertex in the solution set. The stated memory complexities can only be attained by making good use of memory. For example, when branching, we re-use the branch set already in memory, and when returning out of recursion, we select again the subset of memory which (still) corresponds to the memory of that branch. When returning from recursion to continue with another branching option, it might also pose a memory problem to have in memory what these branching options are (with $b$ branching options along $k$ recursion steps this requires $\Ot(bk)$ bits of memory). To overcome this, we can recompute the branching options when returning out of recursion, increasing the time complexity or number of passes by a factor $b$. Such memory complexity tricks are also used in \cite[Appendix B.1]{MinOnesdSAT}.

%% file: ms_adaptingkernels.tex
\section{Adapting Existing Kernels}\label{sec:adaptingkernels}
We first show that very general kernels for vertex cover parameterization admit straightforward adaptations to the \textsc{AL} streaming model. The kernels considered are those by Fomin et al.~\cite{jansenPaper} and by Jansen and Kroon~\cite{JansenKroonNewPiFreeDeletion}.
Fomin~et al.~\cite{jansenPaper} provide general kernelization theorems that make extensive use of a single property, namely that some graph properties can be characterized by few adjacencies.

\begin{definition}{(\cite[Definition~3]{jansenPaper})}\label{def:fewadjacencies}
	A graph property $\Pi$ is \emph{characterized by $c_\Pi \in \N$ adjacencies} if for all graphs $G\in \Pi$, for every vertex $v\in V(G)$, there is a set $D\subseteq V(G)\setminus\{v\}$ of size at most $c_\Pi$ such that all graphs $G'$ that are obtained from $G$ by adding or removing edges between $v$ and vertices in $V(G)\setminus D$, are also contained in $\Pi$.
\end{definition}

Fomin et al. show that graph problems such as \problemPfDV{}, can be solved efficiently through kernelization when $\Pi$ is characterized by few adjacencies (and meets some other demands), by making heavy use of the \textsc{Reduce} algorithm they provide. 
The idea behind the \textsc{Reduce} algorithm is to save \textit{enough} vertices with specific adjacencies in the vertex cover, and those vertices that we forget have equivalent vertices saved to replace them. The sets of adjacencies we have to consider can be reduced by making use of the characterization by few adjacencies, as more than $c_\Pi$ adjacencies are not relevant. The number of vertices we retain is ultimately dependent on $\ell \leq K$. 
The kernel by Fomin et al.~\cite{jansenPaper} is given as follows.

\begin{algorithm}
	\caption{\textsc{Reduce}(Graph $G$, Vertex Cover $X\subseteq V(G)$, $r\in \N$, $c\in \N$) \cite[Algorithm~1]{jansenPaper}}
	\label{reduce}	
\begin{algorithmic}
	\ForAll{$Y \in {X \choose \leq c}$ and a partition of $Y$ into $Y^+ \cup Y^-$}
		\State $Z \coloneqq \{v\in V(G)\setminus X \mid v \text{ is adjacent to all of } Y^+ \text{ and none of } Y^-\}$
		\State Mark $r$ arbitrary vertices of $Z$ (if $|Z| < r$ then mark all of them)
	\EndFor
	\State Delete from $G$ all unmarked vertices that are not contained in $X$
\end{algorithmic}
\end{algorithm}

In the AL streaming model, we have enough information to compute this kernel, by careful memory management in counting adjacencies towards specific subsets of the vertex cover. We now go into detail on the streaming adaptation of this kernel, as given in Algorithm~\ref{reducestr}.

\begin{theorem}
    For a fixed constant $c_\Pi$, \textsc{ReduceStr}($G$, $X$, $r$, $c_\Pi$), where $G$ is provided as a stream in the \textsc{AL} model, is a streaming equivalent of \textsc{Reduce}, using one pass, and $\O((|X| + |X|^{c_\Pi}) \log(n))$ bits of memory resulting in a kernel on $\O(|X| + r \cdot |X|^{c_\Pi})$ vertices, output as an \textsc{EA} stream.\label{ThmReduceStr}
\end{theorem}
\begin{proof}
        Let us first elaborate on the working of Algorithm~\ref{reducestr}. The set $Z$ stores all the partitions considered in the original \textsc{Reduce}, together with two counters per partition. The first counter, $x$, tracks the total amount of vertices already `marked' with this partition, which may not exceed $r$. The second counter is reset at every vertex, and tracks whether $v$ is adjacent to the entirety of $Y^+$. This means these two counters mimic exactly the marking behaviour that \textsc{Reduce} applies, except that the marking is not on arbitrary vertices, but dependent on the order of the stream (this does not impact the correctness). The rest of the algorithm merely interacts with $Z$ correctly and uses some storage, $S$ and $V'$, to make sure the output is constructed correctly without using too much memory. $V'$ remembers which vertices in $X$ we have already seen, to avoid outputting the same edge twice. $S$ saves the set of edges adjacent to a vertex $v$, and if we mark $v$, we output $S$.
        
        Let it be clear by the above motivation that the output of \textsc{ReduceStr} can also be an output of \textsc{Reduce}, and therefore, the algorithm works correctly. Let us analyse the space usage. The main concern is the space usage of the set $Z$, which contains partitions of sets in ${X \choose \leq c}$. There are at most $|X|^c$ such sets, each with at most $2^c$ partitions, and each set has at most $c$ elements (using $\log n$ space). This means $Z$ uses $c \cdot 2^c \cdot |X|^c \cdot \log n = \O(|X|^c \log n)$ bits of space. The set $S$ is reset at every vertex $v$ (which is not part of the vertex cover), and contains at most all edges incident on $v$. As $v$ is not an element of the vertex cover, the degree of $v$ is at most $|X|$. So the set $S$ uses at most $\O(|X|\log n)$ bits of space. The set $V'$ has size at most $|X|$, and so uses at most $\O(|X|\log n)$ bits of space. All in all, our memory never exceeds $\O((|X| + |X|^{c}) \log n)$ bits.
\end{proof}

\begin{algorithm}[ht]
	\caption{\textsc{ReduceStr}(Graph $G = (V,E)$ given as a stream in the \textsc{AL} model, Vertex Cover $X\subseteq V(G)$, $r\in \N$, $c\in \N$)}
	\label{reducestr}
	\begin{algorithmic}[1]
		\ForEach{$Y \in {X \choose \leq c}$ and a partition of $Y$ into $Y^+ \cup Y^-$} \Comment{Calculate and store partitions}
			\State Store $(Y^+,Y^-, 0, 0)$ in $Z$
		\EndFor
		\State Store the output vertices $V' \gets \emptyset$ \Comment{Required for neatly outputting $X$}
		\ForEach{$v \in V$ in the stream} \Comment{This entire loop requires only one pass}
			\If{$v\in X$}
				\ForEach{$(v,w) \in E$ in the stream} 
					\If{$w\in V'$} Output $(v,w)$ as part of the kernel \EndIf 
				\EndFor
				\State $V' \gets V' \cup \{v\}$
			\Else \Comment{$v\notin X$}
			\ForEach{$(Y^+, Y^-, x, y) \in Z$} \Comment{Reset local counters}
				\State $y \gets 0$
			\EndFor
			\State Store an edge set $S \gets \emptyset$ \Comment{Reset local edge memory}
			\ForEach{$(v,w) \in E$ in the stream}
				\ForEach{$(Y^+, Y^-, x, y) \in Z$ where $x\leq r$ and $y \geq 0$} \Comment{Count `correct' partitions}
					\If{$w\in Y^+$} $y \gets y+1$ \EndIf
					\If{$w\in Y^-$} $y \gets -1$ \EndIf
				\EndFor
				\State $S \gets S \cup \{(v,w)\}$. \Comment{If we mark $v$, then $(v,w)$ needs to be added}
			\EndFor
			\If{$\exists (Y^+, Y^-, x, y) \in Z$ where $y = |Y^+|$} \Comment{Mark $v$ and output what we can}
				\State $x \gets x+1$ \Comment{Increment $x$ so that this partition marks at most $r$ vertices}
				\State Output $S$ as part of the kernel
			\EndIf
			\EndIf
		\EndFor
	\end{algorithmic}
\end{algorithm}

An interesting observation is that Algorithm~\ref{reducestr} outputs the kernel as an \textsc{EA} stream. This should not have a lot of impact, as we probably want to store the entire kernel anyway. However, if we want the algorithm itself to store the entire kernel, this might increase the memory use, as we would have to store all edges contained in the kernel. The same goes for if we would want to output the kernel as an \textsc{AL} stream (which is how our input is provided). Let us shortly analyse how much memory would be needed for this. For a vertex cover $X$ in a graph $G$, saving $G[X]$ entirely can take up to $\O(|X|^2\log n)$ bits. Next to the vertex cover, the output kernel consists of $\O(r \cdot |X|^c)$ vertices, each of degree at most $|X|$, as these vertices are not part of the vertex cover. Therefore, the total memory use with this approach can be $\O((|X|^2 + r\cdot |X|^{c+1})\log n)$ bits.

The following theorem then shows how this algorithm leads to streaming kernels for \problemPfDV{} as an adaptation of \cite[Theorem~2]{jansenPaper}. We call a graph $G$ \emph{vertex-minimal} with respect to $\Pi$ if $G\in \Pi$ and for all $S \subsetneq V(G)$, $G[S] \notin \Pi$.

\begin{theorem}
    If $\Pi$ is a graph property such that:
	\begin{enumerate}[noitemsep,label={(\roman*)}]
		\item $\Pi$ is characterized by $c_\Pi$ adjacencies,
		\item every graph in $\Pi$ contains at least one edge, and
		\item there is a non-decreasing polynomial $p: \N \rightarrow \N$ such that all graphs $G$ that are vertex-minimal with respect to $\Pi$ satisfy $|V(G)| \leq p(K)$,
	\end{enumerate}
	then \problemPfDV{} admits a kernel on $\O((K + p(K))K^{c_\Pi})$ vertices in the \textsc{AL} streaming model using one pass and $\O((K + K^{c_\Pi}) \log(n))$ space. \label{TheoremPiFree}
\end{theorem}
\begin{proof}
        Combine Theorem~\ref{ThmReduceStr} with the proof of \cite[Theorem~2]{jansenPaper}, where instead of calling \textsc{Reduce($G$, $X$, $\ell + p(|X|)$, $c_\Pi$)} we call \textsc{ReduceStr($G$, $X$, $\ell + p(|X|)$, $c_\Pi$)}.
\end{proof}

We note that the theorem applies to \textsc{$\mathcal{F}$-Subgraph Deletion [VC]} when $\Delta(\mathcal{F})$ (the maximum degree) is bounded as well as to \textsc{Cluster Vertex Deletion~[VC]}. As such, our streaming kernels generalize the kernels of Bishnu et al. \cite{SaketVCNr} for these problems, while the memory requirements and kernel sizes are fairly comparable. Next we have a discussion and further implications for several general problems, following Fomin et al.~\cite{jansenPaper}.

Let us go over a few applications of Theorem~\ref{TheoremPiFree}. Consider \textsc{Cluster Vertex Deletion [VC]}. This problem is exactly \problemPfDV{} where $\Pi=\{P_3\}$, as any $P_3$-free graph can only contain clusters. Notice that $c_\Pi = 2$ and $p(K) = 3$ suffice to meet the demands of Theorem~\ref{TheoremPiFree}. This implies that, given that we already have a vertex cover for our graph, we have a one-pass streaming algorithm for \textsc{CVD} in the \textsc{AL} model using $\O(K^2 \log n)$ space (or $\O(K^3 \log n)$ bits of space to save the kernel). Considering that this algorithm is a simple adaptation of known results, it is interesting to observe that it compares quite well to e.g. a more result of Bishnu et al.~\cite{SaketVCNr}, who show that \textsc{CVD} admits a one-pass (randomized) streaming algorithm in the \textsc{AL} model using $\O(K^2 \log^4(n))$ space when parameterized by the size $K$ of a vertex cover. Note that the kernel sizes differ, as the kernel given by Bishnu et al.~\cite{SaketVCNr} uses $\O(K^2 \log^2n)$ bits of space, while the kernel from Algorithm~\ref{reducestr} uses $\O(K^3 \log n)$ bits of space.

Next consider \textsc{Triangle Deletion [VC]}. This problem is exactly \problemPfDV{} where $\Pi = \{C_3\}$. Once again, it is easy to observe that $c_\Pi = 2$, and $p(K) = 3$ suffice to meet the demands of Theorem~\ref{TheoremPiFree}. This means that, given that we already have a vertex cover, we have a one-pass streaming algorithm for \textsc{TD} in the \textsc{AL} model using $\O(K^3\log n)$ bits of space (if we save the entire kernel). Once again, this result compares very well to a result of Bishnu et al.~\cite{SaketVCNr}, who show that \textsc{TD} admits a one-pass streaming algorithm in the \textsc{AL} model using $\Ot(K^3)$ bits of space when parameterized by the size $K$ of a vertex cover.

Most interesting is that Theorem~\ref{TheoremPiFree} also applies to \textsc{$\mathcal{F}$-Subgraph Deletion [VC]} when $d = \Delta(\mathcal{F}) \leq K$ is also bounded. Bishnu et al.~\cite{SaketVCNr} showed that this problem admits a one-pass, $\O(d \cdot K^{d+1} \cdot \log n)$ bits memory, algorithm in the \textsc{AL} model. To apply Theorem~\ref{TheoremPiFree} to \textsc{$\mathcal{F}$-Subgraph Deletion [VC]}, construct $\Pi$ from $\mathcal{F}$ by adding for each $F \in \mathcal{F}$ to $\Pi$ both $F$ and all graphs $F'$ obtained by adding some set of edges to $F$ (i.e. we end up with all graphs that contain $F$ as a subgraph). Notice that this $\Pi$ is characterized by $c_\Pi = \Delta(\mathcal{F})$ adjacencies, and because $\Delta(\mathcal{F}) \leq K$, the size of every graph is also bounded by $p(K) = \O(K^2)$. Now Theorem~\ref{TheoremPiFree} gives us a kernel of $\O(K^{\Delta(\mathcal{F})+2})$ vertices obtained in one pass and $\O(K^{\Delta(\mathcal{F})}\log n)$ space in the AL model. This implies a one-pass algorithm using $\O(K^{\Delta(\mathcal{F})+2}\log n)$ space for the \textsc{$\mathcal{F}$-Subgraph Deletion [VC]} problem when $\Delta(\mathcal{F}) \leq K$ in the AL model. Notice that this result is comparable to that of Bishnu et al.~\cite{SaketVCNr}, while our kernel is more generally applicable.

The second general result is for finding largest induced subgraphs.

\begin{problemParam}
	\problemtitle{\textsc{Largest Induced $\Pi$-Subgraph [\textsc{VC}]}}
	\probleminput{A graph $G$ with a vertex cover $X$, and an integer $\ell \geq 1$.}
	\problemparameter{The size $K \coloneqq |X|$ of the vertex cover.}
	\problemquestion{Is there a set $P\subseteq V(G)$ of size at least $\ell$ such that $G[P] \in \Pi$?}
\end{problemParam}

With this definition, we give the following theorem as an adaptation of \cite[Theorem~3]{jansenPaper}.

\begin{theorem}
    If $\Pi$ is a graph property such that:
	\begin{enumerate}[noitemsep,label={(\roman*)}]
		\item $\Pi$ is characterized by $c_\Pi$ adjacencies, and
		\item there is a non-decreasing polynomial $p: \N \rightarrow \N$ such that all graphs $G \in \Pi$ satisfy $|V(G)| \leq p(K)$,
	\end{enumerate}
	then \textsc{Largest Induced $\Pi$-Subgraph [\textsc{VC}]} admits a kernel on $\O(p(K) \cdot K^{c_\Pi})$ vertices in the \textsc{AL} streaming model using one pass and $\O((K + K^{c_\Pi}) \log(n))$ bits of space.\label{thm:LargestPiSubgraph}
\end{theorem}
\begin{proof}
        Combine Theorem~\ref{ThmReduceStr} with the proof of \cite[Theorem~3]{jansenPaper}, where instead of calling \textsc{Reduce($G$, $X$, $p(|X|)$, $c_\Pi$)} we call \textsc{ReduceStr($G$, $X$, $p(|X|)$, $c_\Pi$)}.
\end{proof}

Examples of problems fitting in the \textsc{Largest Induced $\Pi$-Subgraph [VC]} category, that are also characterized by few adjacencies, are \textsc{Long Cycle}, \textsc{Long Path}, and \textsc{$H$-Packing}.

The third general result is for graph partitioning problems.

\begin{problemParam}
	\problemtitle{\textsc{Partition into $q$ Disjoint $\Pi$-Free Subgraphs~[\textsc{VC}]}}
	\probleminput{A graph $G$ with a vertex cover $X$.}
	\problemparameter{The size $K \coloneqq |X|$ of the vertex cover.}
	\problemquestion{Is there a partition of the vertex set into $q$ sets $S_1 \cup S_2 \cup \ldots \cup S_q$ such that for each $i\in [q]$ the graph $G[S_i]$ does not contain a graph in $\Pi$ as an induced subgraph?}
\end{problemParam}

Note that $q$ is regarded a constant.

\begin{theorem}
    If $\Pi$ is a graph property such that:
	\begin{enumerate}[noitemsep,label={(\roman*)}]
		\item $\Pi$ is characterized by $c_\Pi$ adjacencies, and
		\item there is a non-decreasing polynomial $p: \N \rightarrow \N$ such that all graphs such that all graphs $G$ that are vertex-minimal with respect to $\Pi$ satisfy $|V(G)| \leq p(K)$,
	\end{enumerate}
	then \textsc{Partition into $q$ Disjoint $\Pi$-Free Subgraphs~[\textsc{VC}]} admits a kernel on $\O(p(K) \cdot K^{q \cdot c_\Pi})$ vertices in the \textsc{AL} streaming model using one pass and $\O((K + K^{q \cdot c_\Pi}) \log(n))$ bits of space.\label{thm:PartitionqPifree}
\end{theorem}
\begin{proof}
        Combine Theorem~\ref{ThmReduceStr} with the proof of \cite[Theorem~4]{jansenPaper} and \cite[Lemma~2]{jansenPaper}, where instead of calling \textsc{Reduce($G$, $X$, $q \cdot p(|X|)$, $q \cdot c_\Pi$)} we call \textsc{ReduceStr($G$, $X$, $q \cdot p(|X|)$, $q \cdot c_\Pi$)}.
\end{proof}

Examples of problems fitting in the \textsc{Partition into $q$ Disjoint $\Pi$-Free Subgraphs~[VC]} category, that are also characterized by few adjacencies, are \textsc{Partition into $q$ Independent Sets}, \textsc{Partition into $q$ Cliques}, \textsc{Partition into $q$ Planar Graphs}, and \textsc{Partition into $q$ Forests}.

\subsection{Kernel for Characterization by Low-Rank Adjacencies}

We also give an adaptation of a more recent kernel by Jansen and Kroon~\cite{JansenKroonNewPiFreeDeletion}, which has another broad range of implications for streaming kernels. This kernel uses a different characterization of the graph family, however, the adaptation to the \textsc{AL} streaming model is very similar.
The adaptation of this kernel leads to streaming algorithms for problems like \textsc{Perfect Deletion~[VC]}, \textsc{AT-free Deletion~[VC]}, \textsc{Interval Deletion~[VC]}, and \textsc{Wheel-free Deletion~[VC]}.

The characterization for $\Pi$ in this case is a characterization by low-rank adjacencies. For this, however, we need the definition of a c-rank incidence vector.

\begin{definition}{(\cite[Definition~5]{JansenKroonNewPiFreeDeletion})}
	Let $G$ be a graph with vertex cover $X$ and let $c\in \N$. Let $Q', R' \subseteq X$ such that $|Q'| + |R'| \leq c$ and $Q' \cap R' =\emptyset$. We define the \emph{c-incidence vector $\text{inc}_{(G,X)}^{c,(Q',R')}(u)$} for a vertex $u\in V(G)\setminus X$ as a vector over $\Ftwo$ that has an entry for each $(Q,R) \subseteq X \times X$ with $Q \cap R = \emptyset$ such that $|Q| + |R| \leq c$, $Q' \subseteq Q$ and $R'\subseteq R$. It is defined as follows:
	\begin{equation*}
		\text{inc}_{(G,X)}^{c,(Q',R')}(u)[Q,R] =
		\begin{cases}
		1 \quad\text{if }N_G(u) \cap Q = \emptyset \text{ and } R \subseteq N_G(u),\\
		0 \quad\text{otherwise.}
		\end{cases}
	\end{equation*}
\end{definition}

The superscript $(Q',R')$ is dropped when $Q'=R'=\emptyset$. Note that the order of the coordinates of the vector is fixed, but not explicit, as any order suffices. Therefore, we can also sum such incidence vectors coordinate-wise.

With this, we can give the defintion of a graph property being characterized by rank-$c$ adjacencies.

\begin{definition}{(\cite[Definition~7]{JansenKroonNewPiFreeDeletion})}
	Let $c \in \N$ be a natural number. A graph property $\Pi$ is characterized by rank-$c$ adjacencies if the following holds. For each graph $H$, for each vertex cover $X$ of $H$, for each set $D\subseteq V(H)\setminus X$, for each $v\in V(H)\setminus (D \cup X)$, if
	\begin{itemize}[noitemsep]
		\item $H - D \in \Pi$, and
		\item $\text{inc}_{(H,X)}^c(v) = \sum_{u\in D} \text{inc}_{(H,X)}^c(u)$ when evaluated over $\Ftwo$,
	\end{itemize}
	then there exists $D'\subseteq D$ such that $H - v - (D\setminus D')\in \Pi$. If there always exists such set $D'$ of size 1, then we say $\Pi$ is characterized by rank-$c$ adjacencies with singleton replacements.
\end{definition}

Jansen and Kroon note that the intuition here is that if we have a set $D$ such that $H - D \in \Pi$, and the $c$-incidence vectors of $D$ sum to the vector of some vertex $v$ over $\Ftwo$, then there exists $D'\subseteq D$ such that removing $v$ from $H - D$ and adding back $D'$ results in a graph that is still contained in $\Pi$. We can notice how there is some similarity in essential and non-essential adjacencies for occurrences of graphs in $\Pi$ when comparing this characterization to that of few adjacencies (see Definition~\ref{def:fewadjacencies}).

We will give the kernelization algorithm as given by Jansen and Kroon~\cite{JansenKroonNewPiFreeDeletion} next, but first, we recall a linear algebraic definition, that of a \emph{basis}. As Jansen and Kroon state~\cite{JansenKroonNewPiFreeDeletion}, ``a \emph{basis} of a set $S$ of $d$-dimensional vectors over a field $\mathbb{F}$ is a minimum-size subset $B\subseteq S$ such that all $\vec{v}\in S$ can be expressed as linear combinations of elements of $B$, i.e., $\vec{v} = \sum_{\vec{u}\in B}\alpha_{\vec{u}}\cdot \vec{u}$ for a suitable choice of coefficients $\alpha_{\vec{u}}\in \mathbb{F}$. When working over the field $\Ftwo$, the only possible coefficients are $0$ and $1$, which gives a basis $B$ of $S$ the stronger property that any vector $\vec{v}\in S$ can be written as $\sum_{\vec{u}\in B'}\vec{u}$, where $B'\subseteq B$ consists of those vectors which get a coefficient of 1 in the linear combination''.

The essence of the kernel comes down to computing the basis of a set of incidence vectors of the remaining graph and adding vertices corresponding to the basis to the kernel. We give this kernel here as Algorithm~\ref{alg:low-rank-reduce}.

\begin{algorithm}
	\caption{\textsc{Low-Rank Reduce}(Graph $G$, Vertex Cover $X\subseteq V(G)$, $\ell \in \N$, $c\in \N$) \cite[Algorithm~1]{JansenKroonNewPiFreeDeletion}}
	\label{alg:low-rank-reduce}	
	\begin{algorithmic}[1]
		\State Let $Y_1 \coloneqq V(G)\setminus X$
		\For{$i \gets 1$ to $\ell$}
			\State Let $V_i = \{ \text{inc}_{(G,X)}^c(y) \mid y \in Y_i \}$ and compute a basis $B_i$ of $V_i$ over $\Ftwo$.
			\State For each $\vec{v}\in B_i$, choose a unique vertex $y_{\vec{v}} \in Y_i$ such that $\vec{v} = \text{inc}_{(G,X)}^c(y_{\vec{v}})$.
			\State Let $A_i \coloneqq \{ y_{\vec{v}} \mid \vec{v} \in B_i \}$ and $Y_{i+1} = Y_i\setminus A_i$.
		\EndFor
		\State \Return $G[X \cup \bigcup_{i=1}^\ell A_i]$
	\end{algorithmic}
\end{algorithm}

Jansen and Kroon show that Algorithm~\ref{alg:low-rank-reduce} runs in polynomial time in $\ell$ and the size of $G$ (for a constant $c$), and returns a graph on $\O(|X| + \ell \cdot |X|^c)$ vertices. We will now adapt Algorithm~\ref{alg:low-rank-reduce} to the streaming setting, and conclude it to be a streaming equivalent of \cite[Theorem~9]{JansenKroonNewPiFreeDeletion}.

\subsubsection{Adapting Low-Rank Reduce}

If we want to adapt Algorithm~\ref{alg:low-rank-reduce}, \textsc{Low-Rank Reduce}, to the streaming setting, we are faced with a few challenges. For one, the set $V_i$ consists of $\O(n)$ vectors, so saving this entire set is not desirable. We also have to consider how we can compute the basis of the set $V_i$ if we do not want to save it. Luckily, the incidence vectors are computable from local information combined with the vertex cover, and computing a basis can be done incrementally by checking linear (in)dependence. With this small motivation, we give the adaptation of \textsc{Low-Rank Reduce} into the streaming setting, \textsc{Low-Rank ReduceStr}, in Algorithm~\ref{alg:low-rank-reduce-stream}.

\begin{algorithm}
	\caption{\textsc{Low-Rank ReduceStr}(Graph $G$ as an AL stream, Vertex Cover $X\subseteq V(G)$, $\ell \in \N$, $c\in \N$)}
	\label{alg:low-rank-reduce-stream}	
	\begin{algorithmic}[1]
		\State Let $A \gets \emptyset$
		\For{$i \gets 1$ to $\ell$}
		\State $A_i \gets \emptyset$
		\State $B \gets \emptyset$
		\ForEach{Vertex $v\in V \setminus (X \cup A)$ in the stream} \Comment{Entire loop in one pass}
		\State Save the $\leq |X|$ adjacencies of $v$ in $X$ (until the next vertex)
		\State Let $\vec{v} \gets \text{inc}_{(G,X)}^c(v)$ \Comment{Can be computed from $X$ and the adjacencies of $v$}
		\State If $\vec{v}$ is linearly independent w.r.t. $B$ over $\Ftwo$, do $B \gets B \cup \{\vec{v}\}$ and $A_i \gets A_i \cup \{v\}$.
		\EndFor
		\State Let $A \gets A \cup A_i$.
		\EndFor
		\State \Return $G[X \cup A]$
	\end{algorithmic}
\end{algorithm}

\begin{theorem}
    \label{thm:alg-low-rank-reduce-stream}
	Algorithm~\ref{alg:low-rank-reduce-stream} is a streaming equivalent of Algorithm~\ref{alg:low-rank-reduce}, that is, given a graph $G$ as an AL stream with a vertex cover $X$, and integer $\ell$ and a constant $c$, Algorithm~\ref{alg:low-rank-reduce-stream} returns a graph on $\O(|X| + \ell \cdot |X|^c)$ vertices as an AL stream that could be the output of Algorithm~\ref{alg:low-rank-reduce} given the same input. Algorithm~\ref{alg:low-rank-reduce-stream} uses $\ell + 1$ passes and $\O((|X| + \ell \cdot |X|^c)\log n + |X|^{2c})$ bits of memory.
\end{theorem}
\begin{proof}
        Let us first show the equivalence of Algorithm~\ref{alg:low-rank-reduce-stream} to Algorithm~\ref{alg:low-rank-reduce}. Let $G$ be a graph with vertex cover $X$, and let $\ell$ and $c$ be integers. Algorithm~\ref{alg:low-rank-reduce-stream} and Algorithm~\ref{alg:low-rank-reduce} both do $\ell$ iterations over some process over all vertices, excluding the vertices in $X$ and a set $A$ (for both algorithms, $A = \bigcup_{i=1}^\ell A_i$). With these vertices, Algorithm~\ref{alg:low-rank-reduce} computes the incidence vector of each of them, and then computes a basis for this set. The new vertices added to $A$ are then vertices with incidence vectors equivalent to those in the basis. Algorithm~\ref{alg:low-rank-reduce-stream} computes the incidence vectors of these vertices one vertex at a time. It then checks whether the current incidence vector is linearly independent of a set $B$, and if so, the incidence vector is added to $B$. We can see that $B$ must consist of a basis of all incidence vectors seen so far. Therefore, after Algorithm~\ref{alg:low-rank-reduce-stream} has seen every vertex (excluding those in $X$ and $A$), $B$ is a basis that could also be found by Algorithm~\ref{alg:low-rank-reduce} for the same set of vertices. We conclude that in every iteration, Algorithm~\ref{alg:low-rank-reduce-stream} adds to $A$ a set of vertices which Algorithm~\ref{alg:low-rank-reduce} can also add to $A$ in the corresponding iteration. As both algorithms return $G[X \cup A]$, the output of Algorithm~\ref{alg:low-rank-reduce-stream} can also be an output of Algorithm~\ref{alg:low-rank-reduce}.
	
	    As the incidence vector of a vertex $v$ can be computed from the adjacencies of $v$ together with $X$ alone, and linear (in)depence checking only requires the vectors to be in memory, we only require one pass for each $1 \leq i \leq \ell$. Another pass is used to compute the output, as we can produce an AL stream corresponding to $G[X \cup A]$ by simply using a pass and output only those edges between vertices in $X \cup A$. Therefore, Algorithm~\ref{alg:low-rank-reduce-stream} uses $\ell + 1$ passes over the stream.
	
	    In terms of memory, Algorithm~\ref{alg:low-rank-reduce-stream} stores $A$, $B$, $X$, adjacencies of a vertex $v$, and an incidence vector of $v$, $\vec{v}$. Clearly, the memory used by the adjacencies of $v$ and $\vec{v}$ is dominated by saving $X$ and $B$ in memory. The computation of $\vec{v}$ also requires no more memory than $X$ uses, as it iterates subsets of $X$ to compute entries of the vector. Saving $X$ requires $\Ot(|X|)$ bits of memory. $B$ consists of at most $\O(|X|^c)$ vectors, and each vector consists of $\O(|X|^c)$ bits, as motivated by Jansen and Kroon~\cite[Proposition~8]{JansenKroonNewPiFreeDeletion}. It follows that $B$ requires $\O(|X|^{2c})$ bits of space. The set $A$ consists of at most $\O(\ell \cdot |X|^c)$ vertices, as in every iteration $B$ contains at most $\O(|X|^c)$ vectors. We conclude that the total memory usage is $\O((|X| + \ell \cdot |X|^c)\log n + |X|^{2c})$ bits.
\end{proof}

With Theorem~\ref{thm:alg-low-rank-reduce-stream} we are ready to give the streaming equivalent of \cite[Theorem~9]{JansenKroonNewPiFreeDeletion}.

\begin{theorem}
    \label{thm:PiFreeDeletion-RankC}
	If $\Pi$ is a graph property such that:
	\begin{enumerate}[noitemsep]
		\item[(i)] $\Pi$ is characterized by rank-$c$ adjacencies,
		\item[(ii)] every graph in $\Pi$ contains at least one edge, and
		\item[(iii)] there is a non-decreasing polynomial $p : \N \rightarrow \N$ such that all graphs $G$ that are vertex-minimal with respect to $\Pi$ satisfy $|V(G)| \leq p(K)$,
	\end{enumerate}
	then \problemPfDV{} in the AL streaming model admits a kernel on $\O((K + p(K)) \cdot K^c)$ vertices using $K + p(K) + 2$ passes and $\O((K + p(K)) \cdot K^c\log n + K^{2c})$ bits of memory.
\end{theorem}
\begin{proof}
        See the proof of \cite[Theorem~9]{JansenKroonNewPiFreeDeletion}, where instead of \textsc{Low-Rank Reduce$(G, X, \ell \coloneqq k + 1 + p(|X|), c)$} we call \textsc{Low-Rank ReduceStr$(G, X, \ell \coloneqq k + 1 + p(|X|), c)$}. By Theorem~\ref{thm:alg-low-rank-reduce-stream} the theorem follows.
\end{proof}

Let us shortly list some implications of Theorem~\ref{thm:PiFreeDeletion-RankC}, which consist of some problems admitting streaming kernels, following Jansen and Kroon~\cite{JansenKroonNewPiFreeDeletion}.

The first result is for \textsc{Perfect Deletion [VC]}. \textsc{Perfect Deletion [VC]} is \problemPfDV{} where $\Pi$ is the set of all graphs that contain an odd hole or an odd anti-hole. An \emph{odd hole} is a cycle consisting of an odd number of vertices, and an \emph{odd anti-hole} is the complement graph of an odd hole.

\begin{theorem}
    \label{thm:PerfectDeletionLowRank}
	\textsc{Perfect Deletion [VC]} in the AL streaming model admits a kernel on $\O(K^5)$ vertices using $\O(K)$ passes and $\O(K^5\log n + K^8)$ bits of memory.
\end{theorem}
\begin{proof}
        See \cite[Theorem~19]{JansenKroonNewPiFreeDeletion}, but instead of applying \cite[Theorem~9]{JansenKroonNewPiFreeDeletion} we apply Theorem~\ref{thm:PiFreeDeletion-RankC}.
\end{proof}

The second result is for \textsc{AT-free Deletion [VC]}. This is \problemPfDV{} where $\Pi$ is the set of all graphs that contain an asteroidal triple. An \emph{asteroidal tiple} is a set of three vertices where every two vertices in the triple are connected by a path that avoids the neighbourhood of the third.

\begin{theorem}
    \label{thm:ATfreeDeletionLowRank}
	\textsc{AT-free Deletion [VC]} in the AL streaming model admits a kernel on $\O(K^9)$ vertices using $\O(K)$ passes and $\O(K^9\log n + K^{16})$ bits of memory.
\end{theorem}
\begin{proof}
        See \cite[Theorem~21]{JansenKroonNewPiFreeDeletion}, but instead of applying \cite[Theorem~9]{JansenKroonNewPiFreeDeletion} we apply Theorem~\ref{thm:PiFreeDeletion-RankC}.
\end{proof}

The third result is for \textsc{Interval Deletion [VC]}. This is \problemPfDV{} where $\Pi$ is the set of all graphs that contain either an asteroidal triple or an induced cycle of length at least 4, or both.

\begin{theorem}
    \label{thm:IntervalDeletionLowRank}
	\textsc{Interval Deletion [VC]} in the AL streaming model admits a kernel on $\O(K^9)$ vertices using $\O(K)$ passes and $\O(K^9\log n + K^{16})$ bits of memory.
\end{theorem}
\begin{proof}
        See \cite[Theorem~22]{JansenKroonNewPiFreeDeletion}, but instead of applying \cite[Theorem~9]{JansenKroonNewPiFreeDeletion} we apply Theorem~\ref{thm:PiFreeDeletion-RankC}.
\end{proof}

The fourth result is for \textsc{Wheel-free Deletion [VC]}. This is \problemPfDV{} where $\Pi$ is the set of all graphs that contain a wheel of size at least 3. A \emph{wheel} of size $n \geq 3$ is a set of $n+1$ vertices, where $n$ vertices form a cycle, and one vertex is connected to all vertices on the cycle (the center of the wheel).

\begin{theorem}
    \label{thm:WheelDeletionLowRank}
	\textsc{Wheel-free Deletion [VC]} in the AL streaming model admits a kernel on $\O(K^5)$ vertices using $\O(K)$ passes and $\O(K^5\log n + K^8)$ bits of memory.
\end{theorem}
\begin{proof}
        See \cite[Theorem~24]{JansenKroonNewPiFreeDeletion}, but instead of applying \cite[Theorem~9]{JansenKroonNewPiFreeDeletion} we apply Theorem~\ref{thm:PiFreeDeletion-RankC}.
\end{proof}

It is interesting to note that the above problems are not characterized by few adjacencies, but are characterized by rank-$c$ adjacencies, therefore requiring Algorithm~\ref{alg:low-rank-reduce-stream} to admit a streaming kernel.

%% file: ms_directfptapproach.tex
\section{A Direct FPT Approach}\label{sec:DirectFPT}
In this section, we give direct FPT streaming algorithms for \problemPfDV{} for the same cases as Theorem~\ref{TheoremPiFree}.
This is motivated by the fact that Chitnis and Cormode~\cite{ChitnisTheory} found a direct FPT algorithm for \textsc{Vertex Cover} using $\O(2^k)$ passes and only $\Ot(k)$ space in contrast to the kernel of Chitnis et al.~\cite{ChitnisEsfandiariSampling} using one pass and $\Ot(k^2)$ space. 
Therefore, we aim to explore the pass/memory trade-off for \problemPfDV{} as well.

\subsection{\texorpdfstring{$P_3$-free Deletion}{P3-free Deletion}}
We start with the scenario where $\Pi = \{P_3\}$, which means we consider the problem \textsc{Cluster Vertex Deletion [VC]}.
The general idea of the algorithm is to branch on what part of the given vertex cover should be in the solution. For managing the branching correctly, we use a black box enumeration technique also used by Chitnis and Cormode~\cite{ChitnisTheory}.
\begin{definition}{(\cite[Definition~9]{ChitnisTheory})}\label{def:DictOrd}
	Let $U = \{u_1,u_2,\ldots,u_n\}$ and $k\leq n$. Let $\U_{\leq k}$ denote the set of all $\sum_{i=0}^{k} {|U| \choose i}$ subsets of $U$ which have at most $k$ elements, and let $\textsc{Dict}_{\U_{\leq k}}$ be the dictionary ordering on $\U_{\leq k}$. Given a subset $X\in \U_{\leq k}$, let $\textsc{Dict}_{\U_{\leq k}}(\textsc{Next}(X))$ denote the subset that comes immediately after $X$ in the ordering $\textsc{Dict}_{\U_{\leq k}}$. We denote the last subset in the dictionary order of $\U_{\leq k}$ by $\textsc{Last}(\U_{\leq k})$, and similarly the first subset as $\textsc{First}(\U_{\leq k})$, and use the notation that $\textsc{Dict}_{\U_{\leq k}}(\textsc{Next}(\textsc{Last}(\U_{\leq k}))) = \spadesuit$. Similarly, we define $\U_k$ as the set of all ${|U| \choose k}$ subsets of $U$ with exactly $k$ elements, and analogously define the dictionary ordering on this set.
\end{definition}

\begin{figure}[tb]
	\centering
	\includegraphics[width=\textwidth]{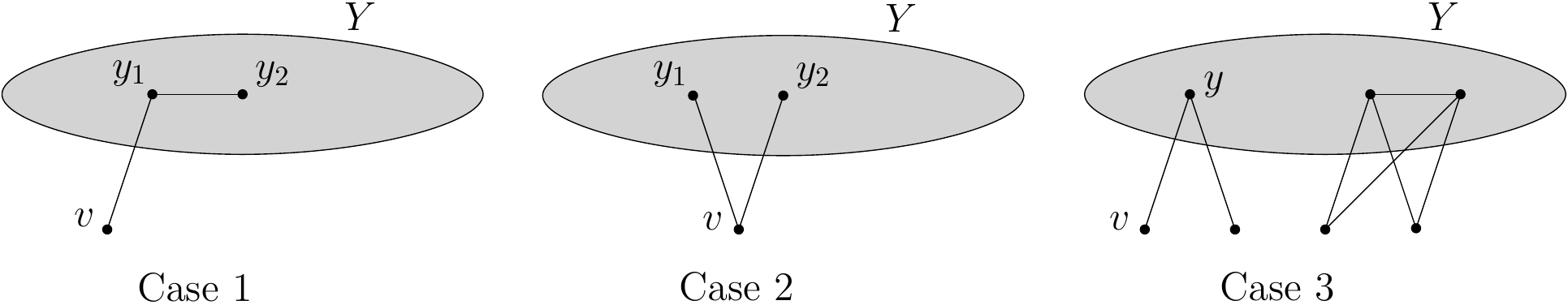}
	\caption{The different cases how a $P_3$ can exists with respect to $Y$, part of the vertex cover. Notice that the case where the entire $P_3$ is contained in $Y$ is not included here. Case 3 assumes there are no Case 1 or Case 2 $P_3$'s in the graph anymore.}
	\label{p3cases}
\end{figure}

In a branch, we first check whether the `deletion-free' part of the vertex cover ($Y$) contains a $P_3$, which invalidates a branch. Otherwise, what remains is some case analysis where either one or two vertices of a $P_3$ lie outside the vertex cover, for which we deterministically know which vertices have to be removed to make the graph $P_3$-free. We illustrate this step in Figure~\ref{p3cases}. Case~1 and~2 have only one option for removal of a vertex. After Case~1 and~2 no longer occur, we can find Case~3 occurrences and show that we can delete all but one of the vertices in such an occurrence. 
So, if this process can be executed in a limited number of passes, the algorithm works correctly.

We give the full algorithm in Algorithm~\ref{alg:p3free}.

\begin{algorithm}[ht]
	\caption{\textsc{$P_3$-free Deletion}(Graph $G = (V,E)$ given as a stream in the \textsc{AL} model, integer $\ell$, Vertex Cover $X\subseteq V(G)$)}
	\label{alg:p3free}
	\begin{algorithmic}[1]
		\State $S \gets \textsc{First}(\X_{\leq \ell})$
		\While{$S \in \X_{\leq \ell}, S \neq \spadesuit$}
			\State $Y \gets X\setminus S$ \Comment{$Y$ is the part of the vertex cover not in the solution $S$}
			\State $S' \gets S$ \Comment{If $S'$ ever exceeds size $\ell$, move to the next $S$}
			\State $P \gets \textsc{First}(\Y_2)$
			\While{$P = (y_1, y_2) \in \Y_2, P \neq \spadesuit$} \Comment{We enumerate all pairs in $Y$}
				\ForEach{Vertex $v \in Y \setminus P$} \State If $v$ and $P$ form a $P_3$, $Y$ is invalid, move to the next $S$ \Comment{Requires a pass}\EndFor
				\State $P \gets \textsc{Dict}_{\Y_2}(\textsc{Next}(P))$
			\EndWhile
			\State $P \gets \textsc{First}(\Y_2)$
			\While{$P = (y_1, y_2) \in \Y_2, P \neq \spadesuit$}\label{line:pairsinVC} \Comment{We enumerate all pairs in $Y$}
				\If{$y_1y_2$ is an edge}
					\ForEach{Vertex $v\in V \setminus (X \cup S')$ in the stream}
						\If{Either $vy_1$ or $vy_2$ is present and the other is not} $S' \gets S' \cup \{v\}$\EndIf
					\EndFor
				\Else
					\ForEach{Vertex $v\in V \setminus (X \cup S')$ in the stream}
						\If{Both $vy_1$ and $vy_2$ are present} $S' \gets S' \cup \{v\}$ \EndIf
					\EndFor
				\EndIf
				\State $P \gets \textsc{Dict}_{\Y_2}(\textsc{Next}(P))$
			\EndWhile
			\ForEach{$y \in Y$}
				\State $b \gets False$
				\ForEach{Vertex $v\in V \setminus (X \cup S')$ in the stream}
					\If{The edge $vy$ is present and $b = False$} $b \gets True$
					\ElsIf{The edge $vy$ is present} $S' \gets S' \cup \{v\}$
					\EndIf
				\EndFor
			\EndFor 
			\If{$|S'| \leq \ell$} \Return $S'$ \EndIf
			\State $S \gets \textsc{Dict}_{\X_{\leq \ell}}(\textsc{Next}(S))$
		\EndWhile
		\State \Return \textsc{NO} \Comment{No branch resulted in a solution}
	\end{algorithmic}
\end{algorithm}

To limit the number of passes, the use of the \textsc{AL} model is crucial. Notice that for every pair of vertices $y_1,y_2$ in the vertex cover, we can identify a Case~1 or~2 $P_3$ of Figure~\ref{p3cases}, or these cases but with $v$ in the vertex cover as well, in a constant number of passes. This is because we can first use a pass to check the presence of an edge between $y_1$ and $y_2$, and afterwards use a pass to check the edges of every other vertex towards $y_1$ and $y_2$ (which are given together because of the \textsc{AL} model). This means we can find $P_3$'s contained in the vertex cover or corresponding to Case~1 or~2 $P_3$'s in $\O(K^2)$ passes total. The remaining Case~3 can be handled in $\O(K)$ passes from the viewpoint of each $y \in Y$. So this algorithm takes $\O(2^K K^2)$ passes (including branching).

\begin{theorem}\label{thm:P3deletion}
    We can solve \textsc{Cluster~Vertex~Deletion~[VC]} in the \textsc{AL} streaming model using $\O(2^KK^2)$ passes and $\O(K \log n)$ space.
\end{theorem}
\begin{proof}
        We claim that Algorithm~\ref{alg:p3free} does exactly this.

        Let us first reason that the number of passes and memory use are as stated. Let $X$ be the provided vertex cover, and let $|X| = K$. The number of different sets $S$ can take is bounded by $2^K$. The first and second loop enumerate all pairs of vertices in the vertex cover, and use a single pass per pair to detect a $P_3$, which gives us at most $K^2$ passes. The last loop enumerates all $K$ vertices in the vertex cover and uses one pass per iteration. Therefore, the total number of passes is bounded by $\O(2^KK^2)$.
        
        No set in memory exceeds $\O(K \log n)$ bits, so the stated memory complexity is correct.
        
        Let us now show the correctness of the algorithm. The main idea of the algorithm is to branch on what part of the vertex cover is contained in the solution $S'$. This is modelled through the use of the sets $Y$ and $S$, where in each branch, we cannot add vertices in $Y$ to $S'$. Therefore, we first check whether $Y$ fully contains a $P_3$, and if one is found we stop, as we may not delete any vertex of this $P_3$ in this branch. For a fixed pair $(v,w)$ in the vertex cover, checking for a $P_3$ that contains $v$ and $w$ only takes one pass because the only necessary information is the adjacencies of $v$ and $w$ towards another vertex, which is provided in the stream local to that vertex (see also Case 1 and Case 2 in the following analysis).
        
        What remains is a careful analysis of the different cases of the structure of $P_3$'s with respect to $Y$. An illustration is given in Figure~\ref{p3cases}. The loop of line~\ref{line:pairsinVC} considers all pairs of vertices in $Y$. There are two cases we are interested in: Case 1 and Case 2 in Figure~\ref{p3cases}. If we look at a single pair of vertices $y_1$ and $y_2$ either there is an edge between them (Case 1) or a non-edge (Case 2). These two vertices can then form a $P_3$ with any vertex outside $Y$ in a very specific manner, which the algorithm looks for. It is then trivial that the one vertex outside $Y$ has to be removed to make the graph $P_3$ free if a $P_3$ is found.
        
        If there are no Case 1 or Case 2 $P_3$'s in the graph any more, we move on to Case 3. Note that this is the only remaining way a $P_3$ can be in the graph at all, because $Y \subseteq X$ is (part of) a vertex cover. In Case 3 at first it seems undecided which of the two vertices outside $Y$ to remove, as one might lead to a solution and the other not. Let $y_1, v, w$ form a Case 3 $P_3$, where $y_1\in Y$. Let us consider the scenario where $v$ has another adjacency $y_2 \in Y$. Because there are no Case 2 $P_3$'s, $y_1$ and $y_2$ must be adjacent. Because there are no Case 1 $P_3$'s, $w$ must now also be adjacent to $y_2$. This means the structure extends as illustrated on the right in Case 3 in Figure~\ref{p3cases}. We can observe that we need to delete all but one of the vertices attached to $y$, which is what the algorithm does. It does not matter which vertex we do not delete, as this vertex forms triangles if it has multiple adjacencies. Therefore, after these cases have all been handled, no induced $P_3$'s remain in the graph. If during the process $S'$ never exceeded size $\ell$, this means we have found a solution; otherwise, we move on to the next branch.
        
        By the above reasoning, if there exists a solution of size at most $\ell$ for the \textsc{Cluster Vertex Deletion [VC]} problem, then this solution contains some subset of the vertex cover $X$, which corresponds to some branch in the algorithm. As the removal of vertices is deterministic  in each branch (as in, the solution must contain these vertices), and there exists a solution, the algorithm must find a solution too in that branch. If there exists no solution of size at most $\ell$, then there exists no subset of vertices $S'$ such that $G \setminus S'$ is induced $P_3$ free, and so in each branch of the algorithm $S'$ will exceed size $\ell$ at some point, which results in the algorithm returning NO.
\end{proof}

Let us stress some details. The use of the \textsc{AL} model is crucial, as it allows us to locally inspect the neighbourhood of a vertex when it appears in the stream. The same approach would require more memory or more passes in other models to accomplish this result.
Also note that we could implement this algorithm in a normal setting (the graph is in memory, and not a stream) to get an algorithm for \textsc{Cluster Vertex Deletion [VC]} with a running time of $\O(2^K \cdot K^2 \cdot (n+m))$.

\subsection{\texorpdfstring{$H$-free Deletion}{H-free Deletion}}\label{sec:HfreeDeletion}

We now consider a more generalized form of \problemPfDV{}, where $\Pi = \{H\}$, a single graph. Unfortunately, the approach when $H=P_3$ does not seem to carry over to this case, because the structure of a $P_3$ is simple and local.

\begin{theorem}\label{thm:generic-upper-fpt}
We can solve \textsc{$H$-free Deletion~[VC]} in $2^{O(K^2)}\, \mathrm{poly}(n,|V(H)|)$ time, where $H$ contains at least one edge and $K$ is the size of the vertex cover.
\end{theorem}
\begin{proof}
Let $X$ be a vertex cover of $G=(V,E)$ of size $K$. Then $G[V \setminus X]$ has no edges and thus does not contain an occurrence of $H$. It follows that there is a solution of size at most $K$. Now call two vertices of $V \setminus X$ equivalent if their neighborhood in $X$ is the same. This yields $2^K$ equivalence classes. Observe that vertices in an equivalence class are interchangeable with respect to a solution for \textsc{$H$-free Deletion [VC]}: one can be exchanged for another without changing the validity of the solution. Hence, we may select the vertices of a solution from the first $K$ vertices of an equivalence class. This means that there is a set of at most $(2^K+1)K$ vertices in $G$ that form a superset of some solution. Then we can enumerate all possible such solutions, which have size at most $K$, in $2^{O(K^2)}$ time. The validity of a solution can be checked in $2^{O(K \log K)} \mathrm{poly}(n,|V(H)|)$ time through the algorithm of Abu-Khzam~\cite{Abu-Khzam14}.
\end{proof}

In order to analyze the complexity with respect to $H$ more precisely and to obtain a streaming algorithm, we present a different algorithm that works off a simple idea.
We branch on the vertex cover, and then try to find occurrences of $H$ of which we have to remove a vertex outside the vertex cover. We branch on these removals as well, and repeat this find-and-branch procedure.
In an attempt to keep the second branching complexity low, we start by searching for occurrences of $H$ such that only one vertex lies outside the vertex cover, and increase this number as we find no occurrences. For clarity, we present the occurrence detection part of the algorithm first, a procedure we call \textsc{FindH}. Note that this is not (yet) a streaming algorithm.

\begin{algorithm}[tb]
	\caption{The procedure \textsc{FindH}.}
	\label{alg:FindH}
	\begin{algorithmic}[1]
		\Function{FindH}{solution set $S$, forbidden set $Y \subseteq X$, integer $i$}
			\ForEach{Set $O$ of $i$ vertices of $H$ that can be outside $X$} \Comment{Check non-edges}
				\State Denote $H' = H \setminus O$
				\ForEach{Occurrence of $H'$ in $Y$} \Comment{Check $\O({|X| \choose |H|-i} (|H|-i)!)$ options}
					\State $S'\gets \emptyset$, $O' \gets O$
					\ForEach{Vertex $v\in V \setminus (S \cup X)$}
						\State Check the edges/non-edges towards $H' \in Y$
						\If{$v$ is equivalent to some $w \in O'$ for $H'$}
							\State $S'\gets S' \cup \{v\}$, $O' \gets O' \setminus \{w\}$
							\If{$O' = \emptyset$} \Return $S'$ \Comment{We found an occurrence of $H$} \EndIf 
						\EndIf
					\EndFor
				\EndFor
			\EndFor
			\State \Return $\emptyset$ \Comment{No occurrence of $H$ found}
		\EndFunction
	\end{algorithmic}
\end{algorithm}

\begin{lemma}
    Given a graph $G$ with vertex cover $X$, graph $H$ with at least one edge, and sets $S$, $Y \subseteq X$, and integer $i$, Algorithm~\ref{alg:FindH} finds an occurrence of $H$ in $G$ that contains no vertices in $S$ and $X \setminus Y$ and contains $|V(H)| - i$ vertices in $Y$. It runs in $\O\left({h \choose i} [i^2 + {K \choose h-i} (h-i)!((h-i)^2 + Kn + (h-i)in)]\right)$ time, where $|V(H)| = h$ and $|X| = K$.\label{lemma:FindH_FPT}
\end{lemma}
\begin{proof}
        The correctness of the algorithm follows from the enumeration of all possibilities.
        
        Let us analyse the running time. Checking all possible sets $O$ takes $\O({h \choose i}i^2)$ time resulting in at most $\O({h \choose i})$ options for $O$. There are at most $\O({K \choose h-i} (h-i)!)$ options for $H'$ in $Y$: checking all of them costs $\O({K \choose (h-i)} (h-i)! (h-i)^2)$ time. Then we take $\O((K + (h - i)i)n)$ time to, for each vertex, save adjacencies to $H'$, and check whether it matches on of those in $O$. The $(h-i)$ factor is for checking adjacencies towards $H'$. Therefore, the running time of \textsc{FindH} is $\O\left({h \choose i} [i^2 + {K \choose h-i} (h-i)!((h-i)^2 + Kn + (h-i)in)]\right)$.
\end{proof}

Now let us give the complete FPT algorithm for \textsc{$H$-free Deletion [VC]} (not in the streaming setting), see Algorithm~\ref{alg:hfreefpt}.

\begin{algorithm}[!htp]
	\caption{\textsc{$H$-free Deletion FPT}(Graph $G = (V,E)$, integer $\ell$, Vertex Cover $X\subseteq V(G)$)}
	\label{alg:hfreefpt}
	\begin{algorithmic}[1]
		\ForEach{Partition of $X$ into $S,Y$ where $|S| \leq \ell$}
			\If{$H$ is not contained in $Y$} \Comment{Check all $\O({|X| \choose |H|} |H|!)$ options}
				\If{\Call{Branch}{$S$, $Y$, $1$}} \State \Return YES \Comment{If any returns YES, we also return YES}\EndIf
			\EndIf
		\EndFor
		\State \Return NO
		
		\Statex
		
		\Function{Branch}{solution set $S$, forbidden set $Y \subseteq X$, integer $i$}
			\State $B \gets$ \Call{FindH}{$S$, $Y$, $i$} \Comment{Try to find an $H$ with $i$ vertices outside $Y$}
			\If{$B = \emptyset$ and $i = |H|$}
				\Return YES
			\ElsIf{$B = \emptyset$}
				\Call{Branch}{$S$, $Y$, $i + 1$} \Comment{No $H$ found}
			\ElsIf{$|S| = \ell$}
				\Return NO \Comment{Found an $H$ but cannot remove it}
			\Else
				\ForEach{$v \in B$}
					\If {\Call{Branch}{$S \cup \{v\}$, $Y$, $i$}} \Return YES\EndIf
				\EndFor
			\EndIf
		\EndFunction
	\end{algorithmic}
\end{algorithm}

Together with Lemma~\ref{lemma:FindH_FPT} we can analyse the performance of Algorithm~\ref{alg:hfreefpt} in its entirety.

\begin{theorem}
    Algorithm~\ref{alg:hfreefpt} is an FPT algorithm for \textsc{$H$-free Deletion [VC]} using $\O(2^Kh^KK^{h+1}h!h^2n)$ time or alternatively $\O(2^Kh^KK!Kh!h^2n)$ time, where $|V(H)| = h$ and $H$ contains at least one edge.\label{thm:AlgHFreeFpt}
\end{theorem}
\begin{proof}
        Let us first go into detail on the correctness of the algorithm. Assume the algorithm returns YES for some instance $G, H, \ell, X$ where $|X| = K$ and $|V(H)| = h$. The only way the algorithm returns YES, is if in some partition of $X$ into $S$ and $Y$ the \textsc{Branch} function returns YES. The \textsc{Branch} function only returns YES if any recursive call returns YES, or when $B = \emptyset$ and $i = h$. As the latter is the only base case, this must have occurred for this instance. As $i$ starts at $1$ and is only ever incremented, we can conclude that for every $i$ at some point $B = \emptyset$ while $|S| \leq \ell$. The algorithm calls on \textsc{FindH} for every $i$ to find if there is an occurrence of $H$ with $i$ vertices outside of $Y \cup S$ and $|V(H)| - i$ vertices in $Y$. By Lemma~\ref{lemma:FindH_FPT}, \textsc{FindH} correctly finds occurrences of $H$ where $i$ vertices are outside $Y$. As the algorithm returned YES, \textsc{FindH} must have returned an empty set for each $i$ at some point, and so no occurrences of $H$ are present in the graph $G[V \setminus S]$ (otherwise, such an occurrence must have $i$ vertices outside $Y \cup S$ for some $i$). This means that the algorithm is correct in returning YES.
        
        For the other direction, assume that for an instance $G, H, \ell, X$ where $|X| = K$ there exists a smallest set $S_{opt}$ such that $G[V \setminus S_{opt}]$ is $H$-free and $|S_{opt}| \leq \ell$. Then $S_{opt}$ must contain some part of the vertex cover $X$, and as we enumerate all possibilities, the algorithm considers this option. As $G[V \setminus S_{opt}]$ is $H$-free, clearly, for every set $B$ the function \textsc{FindH} finds, at least one vertex in $B$ is also in $S_{opt}$. As we branch on each possibility of the vertices in $B$, the algorithm also considers exactly the option where the set $S$ in the algorithm is a subset of $S_{opt}$. This means there is a branch where the algorithm terminates with $S = S_{opt}$, which means it returns YES as $G[V \setminus S_{opt}]$ is $H$-free. We conclude that the algorithm solves \textsc{$H$-free Deletion} correctly.
        
        Let us analyse the running time of the algorithm. There are $\O(2^K)$ possible partitions of $X$ into $S$ and $Y$. Checking whether $H$ is contained in $Y$ takes $\O({K \choose h} h!h^2)$ time. Because $H$ contains at least one edge, we can assume that $\ell \leq K$, as otherwise $X$ is a trivial solution. The function \textsc{Branch} is called in worst case $\O(h^\ell) = \O(h^K)$ times (branching on at most $h$ vertices each time). \textsc{FindH} is called at most once for every $i$ in every branch. By Lemma~\ref{lemma:FindH_FPT}, \textsc{FindH} runs in $\O\left({h \choose i} [i^2 + {K \choose h-i} (h-i)!((h-i)^2 + Kn + (h-i)in)]\right)$ time. Here is where there is some variance in how we round these complexities, namely when concerning e.g. ${K \choose h}$. This is because we can both say ${K \choose h} = \O(K^h)$, and ${K \choose h} = \O(K!)$. Which of these is a tighter bound comes down to the value of $h$ in comparison to $K$. The total complexity of the algorithm comes down to $${\textstyle\O\left(2^K \left({K \choose h}h!h^2 + h^K \sum_{i=1}^{h} {h \choose i} [i^2 + {K \choose h-i} (h-i)!((h-i)^2 + Kn + (h-i)in)]\right)\right)}$$ time, which we can shorten to either $\O(2^Kh^KK^{h+1}h!h^2n)$ or $\O(2^Kh^KK!Kh!h^2n)$ time.
\end{proof}

Before we go into the translation of Algorithm~\ref{alg:hfreefpt} to the streaming model, let us discuss one of its shortcomings. A large part of the complexity of the algorithm comes from the twofold branching (on the vertex cover and on the possible deletions), but next to this, the function \textsc{FindH} largely contributes to the complexity. We can ask ourselves whether or not this function can be made more efficient. This means we are interested in more efficient induced subgraph finding, which is also called induced subgraph isomorphism. This problem has been studied in the literature, with varying degrees of success. There are a couple of main issues with regard to applying such results to this algorithm. For one, many results focus on specific graph structures, e.g.\ finding $r$-regular induced subgraphs~\cite{RegularSubgraphFind}. These results do not help us as we are interested in general structures. Another problematic factor is the common approach of using matrix multiplication. The issue with matrix multiplication is that it does not translate well to the streaming model, as often matrices require at least super-linear memory. An example of such an algorithm can be found in \cite{FindCountSmallSubgraphsEfficiently}.

\textsc{FindH} is adaptable to the streaming setting, as is the complete algorithm, see Algorithm~\ref{alg:hfreestream}. All the actions \textsc{FindH} takes are local inspection of edges, and many enumeration actions, which lend itself well to usage of the AL streaming model. The number of passes of the streaming version is closely related to the running time of the non-streaming algorithm. This then leads to the full find-and-branch procedure.

It should be clear that the functionality of Algorithm~\ref{alg:hfreestream} is the same as that of Algorithm~\ref{alg:hfreefpt}, but translated to the streaming model using as little memory as possible. Once again we make use of dictionary orderings, see Definition~\ref{def:DictOrd} for the formal definition.

\begin{algorithm}[!htp]
	\caption{\textsc{$H$-free Deletion Stream}(Graph $G = (V,E)$ in the AL model, integer $\ell$, Vertex Cover $X\subseteq V(G)$)}
	\label{alg:hfreestream}
	\begin{algorithmic}[1]
		\State $S \gets \textsc{First}(\X_{\leq \ell})$
		\While{$S \in \X_{\leq \ell}, S \neq \spadesuit$}
			\State $Y \gets X\setminus S$
			\State $S' \gets S$
			\If{$\neg$ \Call{Check}{$H$,$Y$}}
				\If{\Call{Branch}{$S'$, $Y$, $1$}} \State \Return YES \Comment{If any returns YES, we also return YES}\EndIf
			\EndIf
			\State $S \gets \textsc{Dict}_{\X_{\leq \ell}}(\textsc{Next}(S))$
		\EndWhile
		\State \Return NO
		
		\Statex
		
		\Function{Check}{set to find $H$, search space $Y$}
			\State $P \gets \textsc{First}(\Y_{|H|})$
			\While{$P \in \Y_{|H|}, P \neq \spadesuit$}
				\ForEach{Permutation $p$ of the vertices of $H$}
					\State Use a pass to check if $p$ matches $P$ \Comment{Go to the next $p$ if some edges do not match}
					\If{$p$ matches $P$} \Return YES \EndIf
				\EndFor
				\State $P \gets \textsc{Dict}_{\Y_{|H|}}(\textsc{Next}(P))$
			\EndWhile
			\State \Return NO
		\EndFunction
		
		\Statex
		
		\Function{Branch}{solution set $S$, forbidden set $Y \subseteq X$, integer $i$}
			\State $B \gets$ \Call{FindH}{$S$, $Y$, $i$} \Comment{Try to find an $H$ with $i$ vertices outside $Y$}
			\If{$B = \emptyset$ and $i = |H|$}
				\Return YES
			\ElsIf{$B = \emptyset$}
				\Call{Branch}{$S$, $Y$, $i + 1$} \Comment{No $H$ found}
			\ElsIf{$|S| = \ell$}
				\Return NO \Comment{Found an $H$ but cannot remove it}
			\Else
				\ForEach{$v \in B$}
					\If {\Call{Branch}{$S \cup \{v\}$, $Y$, $i$}} \Return YES \EndIf
				\EndFor
			\EndIf
		\EndFunction
		
		\Statex
		
		\Function{FindH}{solution set $S$, forbidden set $Y \subseteq X$, integer $i$}
			\ForEach{Set $O$ of $i$ vertices of $H$ that can be outside $Y$} \Comment{Check non-edges in $H$}
				\State Denote $H' = H \setminus O$
				\State $P \gets \textsc{First}(\Y_{|H'|})$
					\While{$P \in \Y_{|H'|}, P \neq \spadesuit$}
						\ForEach{Permutation $p$ of the vertices of $H'$}
							\State Use a pass to check if $p$ matches $P$
							\If{$p$ matches $P$}
								\State $S'\gets \emptyset$, $O' \gets O$
								\ForEach{Vertex $v\in V \setminus (S \cup Y)$}
									\State Check the edges/non-edges towards $H' \in Y$
									\If{$v$ is equivalent to some $w \in O'$ for $H'$}
										\State $S'\gets S' \cup \{v\}$, $O' \gets O' \setminus \{w\}$
										\If{$O' = \emptyset$} \Return $S'$ \Comment{We found an $H$ occurrence} \EndIf 
									\EndIf
								\EndFor
							\EndIf
						\EndFor
						\State $P \gets \textsc{Dict}_{\Y_{|H'|}}(\textsc{Next}(P))$
					\EndWhile
			\EndFor
			\State \Return $\emptyset$ \Comment{No occurrence of $H$ found}
		\EndFunction
	\end{algorithmic}
\end{algorithm}

\begin{theorem}\label{thm:AlgHFreeStream}
    We can solve \textsc{$H$-free Deletion [VC]} in the AL model, where $H$ contains at least one edge, using $\O(2^Kh^{K+2}K^hh!)$ or alternatively $\O(2^Kh^{K+2}K!h!)$ passes and $\O((K + h^2) \log n)$ space, where $|V(H)| = h$.
\end{theorem}
\begin{proof}
    The algorithm in question is Algorithm~\ref{alg:hfreestream}.

    Let it be clear from the algorithm that the approach to solving \textsc{$H$-free Deletion} has not changed from Algorithm~\ref{alg:hfreefpt}. Therefore, if the graph stream is handled correctly, we can conclude that \textsc{$H$-free Deletion} is solved correctly. As we have $H$ in memory, we only require to use passes of the stream to determine (parts of) $X$ and $G$. The dictionary orderings require no passes because we have the vertices of $X$ in memory. The only places where we require passes of the stream thus are when concerning edges of the vertex cover, and edges/vertices in the rest of the graph ($V \setminus X$). Notice that at such points in Algorithm~\ref{alg:hfreestream} we correctly mention the use of a pass. The loop over all vertices in $V \setminus (S \cup Y)$ only requires one pass because of the use of the AL model. Therefore, the algorithm is a correct adaptation of Algorithm~\ref{alg:hfreefpt} to the streaming model.
        
    What remains is to analyse the number of passes and memory use. Let us analyse the memory use per function. The entire algorithm keeps track of the vertex cover $X$, and the forbidden graph $H$. Denote $|X| = K$ and $|V(H)| = h$. The vertex cover uses $\Ot(K)$ bits, and because we save the entirety of $H$ we use $\Ot(h^2)$ bits. The main function uses the set $S$ of at most $\ell$ elements of $X$, likewise $S'$, and the set $Y$ of size $\O(K)$. The function \textsc{Check} uses a set $P$ of $h$ elements, a permutation $p$ of $h$ elements, and $\O(1)$ bits to check if $p$ matches $P$, as it can stop when it does not match. The function \textsc{Branch} uses a set $B$ of at most $h$ elements and increases the size of the set $S$, but only when it does not exceed $\ell$ elements. The function \textsc{FindH} uses sets $O$, $H'$, $P$, $O'$, $p$, all of at most $h$ vertices. It also uses $\O(1)$ bits to check if $p$ matches $P$, and $\O(h)$ bits to save the adjacencies to $H'$ to check if some vertex matches one in $O'$. $S'$ can only contain an element for each element in $O'$. Therefore, $S'$ contains at most $h$ elements.
        
    We can conclude the memory use of a single branch is bounded by $\O((K + h^2)\log n)$ bits. However, in this algorithm we branch on $h$ options, which is not a constant. Therefore, to be able to return out of recursion when branching and continue where we left off, we need to save the set $B$, or recompute it when we return. Saving the sets $B$ takes $\Ot(hK)$ bits because we have at most $K$ active instances. Alternatively, recomputing $B$ adds a factor $h$ to the number of passes. Seeing as we aim to be memory efficient, we opt for this second option here.
        
    The number of passes used by the algorithm is closely aligned with the running time of Algorithm~\ref{alg:hfreefpt}. There are only three places in which we use a pass of the stream, namely, line~14 in the function \textsc{Check}, and line~32 and line~35 in the function \textsc{FindH}. The loop of line 35 requires only one pass because the stream is given in the Adjacency List model. The number of passes is clearly dominated by the number of times the passes in line 32/35 are used. Now we can use the same analysis as for Algorithm~\ref{alg:hfreefpt}, but make some nuances in the running time, as we have to distinguish running time which leads to more passes and running time which will be `hidden' in the allowed unbounded computation. Consider the running time of \textsc{FindH}, as given by Lemma~\ref{lemma:FindH_FPT}, $\O\left({h \choose i} [i^2 + {K \choose h-i} (h-i)!((h-i)^2 + (h-i)in)]\right)$. The running time factors $i^2$ and $(h-i)^2$ are checks over some amount of edges, which will be hidden by the unbounded computation. The factor $(h-i)in$ comes from the finding of vertices that fit the current form of $H$, and can be done in one pass, which means this factor falls away as well. We now have that the \textsc{FindH} function costs us $\O\left({h \choose i} {K \choose h-i} (h-i)!\right)$ passes. This can be shortened to $\O(K^h h!)$ by expanding the ${h \choose i}$ factor, and bounding ${K \choose h-i} = \O(K^h)$. We can also bound this by $\O(K! h!)$ as discussed before. As \textsc{FindH} is called for every $1 \leq i \leq h$ in every branch, we get an extra $h$ factor in the total number of passes. Another factor $h$ is added for recomputing $B$ when returning out of recursion at every step. This means the total number of passes comes down to either $\O(2^Kh^{K+2}K^hh!)$ or $\O(2^Kh^{K+2}K!h!)$.
\end{proof}

\subsection{\texorpdfstring{Towards $\Pi$-free Deletion}{Towards Pi-free Deletion}} \label{subsec:TowardsPifree}
An issue with extending the previous approach to the general \textsc{$\Pi$-free Deletion} problem is the dependence on the maximum size $h$ of the graphs $H \in \Pi$.
Without further analysis, we have no bound on $h$. 
However, we can look to the preconditions used by Fomin et al.~\cite{jansenPaper} on $\Pi$ in e.g.\ Theorem~\ref{TheoremPiFree} to remove this dependence.

The first precondition is that the set $\Pi' \subseteq \Pi$ of graphs that are vertex-minimal with respect to $\Pi$ have size bounded by a function in $K$, the size of the vertex cover. That is, for these graphs $H \in \Pi'$ we have that $|V(H)| \leq p(K)$, where $p(K)$ is some function. We can prove that it suffices to only remove vertex-minimal elements of $\Pi$ to solve \textsc{$\Pi$-free Deletion}, see Lemma~\ref{lemma:vertexMinimalSuffices}. Note that Fomin et al.~\cite{jansenPaper} require that this is a polynomial, we have no need to demand this.

\begin{lemma}
    Let $\Pi$ be some graph property, and denote the set of vertex-minimal graphs in $\Pi$ with $\Pi'$. Let $G$ be some graph and $S \subseteq V(G)$ some vertex set. Then $G[V(G) \setminus S]$ is $\Pi$-free if and only if $G[V(G) \setminus S]$ is $\Pi'$-free.\label{lemma:vertexMinimalSuffices}
\end{lemma}
\begin{proof}
        Assume the preconditions in the lemma, and assume that $G[V(G) \setminus S]$ is not $\Pi'$-free. As $\Pi ' \subseteq \Pi$, clearly, $G[V(G) \setminus S]$ is not $\Pi$-free.
        
        Now assume that $G[V(G) \setminus S]$ is $\Pi'$-free. Assume there is some $H \in \Pi$ such that $H \in G[V(G) \setminus S]$ (that is, $H$ is isomorphic to an induced subgraph of $G[V(G) \setminus S]$). As the graph is $\Pi'$-free, $H$ is a non-vertex-minimal graph with respect to $\Pi$. By definition of vertex-minimal, removal of a specific set of one or more vertices of $H$ results in a vertex-minimal graph $I \in \Pi'$. But if we ignore the same set of vertices in $H \in G[V(G) \setminus S]$, clearly, $I \in G[V(G) \setminus S]$. This contradicts our assumption that $G[V(G) \setminus S]$ is $\Pi'$-free. Therefore, there cannot be a $H \in \Pi$ such that $H \in G[V(G) \setminus S]$, which means that $G[V(G) \setminus S]$ is $\Pi$-free.
\end{proof}

If we also assume that we know the set $\Pi'$, we obtain the following result.

\begin{theorem}\label{thm:polyFPT}
    If $\Pi$ is a graph property such that:
	\begin{enumerate}[noitemsep,label={(\roman*)}]
		\item we have explicit knowledge of $\Pi' \subseteq \Pi$, which is the subset of $q$ graphs that are vertex-minimal with respect to $\Pi$, and
		\item there is a non-decreasing function $p: \N \rightarrow \N$ such that all graphs $G \in \Pi'$ satisfy $|V(G)| \leq p(K)$, and
		\item every graph in $\Pi$ contains at least one edge,
	\end{enumerate}
    then \problemPfDV{} can be solved using $\O(q \cdot 2^K \cdot p(K)^K \cdot K! \cdot K \cdot p(K)! \cdot p(K)^2\cdot n)$ time.
    \label{thm:PiFreeP(K)}
\end{theorem}
\begin{proof}
       This result can be achieved by adjusting Algorithm~\ref{alg:hfreefpt} to search for each of the $q$ graphs in $\Pi'$ instead of only $H$. This increases the complexity by a factor $q$ as we search for more graphs than just one. Then, using Theorem~\ref{thm:AlgHFreeFpt} and Lemma~\ref{lemma:vertexMinimalSuffices}, the theorem follows.
\end{proof}

Note that we require explicit knowledge of $\Pi'$ and $p(K)$ to achieve Theorem~\ref{thm:PiFreeP(K)}. Also note that we chose to write the $K!$ alternative here, as likely $K^{p(K)} > K!$ (i.e. for $p(K) = K$, a linear function, we have that $K^{p(K)} = K^K > K!$).

We argue this algorithm is essentially tight, under the Exponential Time Hypothesis (ETH)~\cite{ImpagliazzoP01}, by augmenting a reduction by Abu-Khzam et al.~\cite{Abu-KhzamBS17}. Recall that the {\problemISI} problem is defined as follows: given two graphs $G_1$ (the host) and $G_2$ (the pattern), does $G_1$ contain an induced subgraph that is isomorphic to $G_2$.
Abu-Khzam et al.~\cite{Abu-KhzamBS17} proved that {\problemISI}[VC] has no $2^{o(K \log K)} \mathrm{poly}(|G_1|,|G_2|)$ time algorithm unless ETH fails, where $K$ is the sum of the vertex cover numbers of $G_1$ and $G_2$. 
We strengthen their reduction to yield the following result.

\begin{theorem}\label{thm:general-lower-fpt}
There is a graph property $\Pi$ for which we cannot solve {\problemPfDV{}} in $2^{o(K \log K)} \mathrm{poly}(n)$ time, unless ETH fails, where $K$ is the vertex cover number of $G$, even if each graph that has property $\Pi$ has size quadratic in its vertex cover number.
\end{theorem}
\begin{proof}
We augment the reduction by Abu-Khzam et al.~\cite{Abu-KhzamBS17}. Their reduction is from the {\problemPC} problem. In this problem, the input is a graph $G=(V,E)$ on the vertex set $V = [k] \times [k]$ and the question is whether it has a clique that contains exactly one vertex from each row and column. That is, whether there exists a set $C \subseteq [k] \times [k]$ such that for each distinct $(i,j), (i',j') \in C$, it holds that $i \not= i'$, $j \not= j'$, and $ij,i'j' \in E$. Then we say that $G$ has a permutation clique of size $k$. Lokshtanov et al.~\cite{LokshtanovMS18} proved that this problem has no $2^{o(k \log k)}$ time algorithm, unless ETH fails.

Abu-Khzam et al.\ show that, given an instance of $(G,k)$ of {\problemPC}, in polynomial time an equivalent instance of {\problemISI}[VC] can be constructed with host graph $G_1$ and pattern graph $G_2$ that both have a vertex cover of size $K = O(k)$. By inspection of their reduction, we observe that $G_2$ has size quadratic in $k$. Moreover, both $G_1$ and $G_2$ have a unique clique (called $D_r$ and $\tilde{D_r}$ respectively) of size~$6$ with designated vertex $r$ and $\tilde{r}$ respectively that is the only vertex that has edges to vertices outside the clique. The isomorphism in the proof of equivalence maps $\tilde{D_r}$ to $D_r$ and $\tilde{r}$ to $r$.

We augment the construction of $G_1$ and $G_2$ as follows. Add a clique of size~$7$ to $G_1$ and $G_2$ (called $A$ and $\tilde{A}$ respectively), with a designated vertex (called $a$ and $\tilde{a}$) respectively. Then add a path $P$ ($\tilde{P}$) of length $k$ from $a$ ($\tilde{a}$) to a designated vertex $d \not= r$ ($\tilde{d} \not= \tilde{r}$). Since $A$ ($\tilde{A}$) is the unique clique of size~$7$ in $G_1$ ($G_2$), any isomorphism from the augmented versions of $G_2$ to $G_1$ will map $\tilde{A}$ to $A$. Since $a$ is the only vertex of $A$ of degree $7$, $\tilde{a}$ will be mapped to $a$. Repeating such arguments, $\tilde{P}$ will be mapped to $P$, $\tilde{d}$ to $d$, $\tilde{D_r}$ to $D_r$, and $\tilde{r}$ to $r$. Then the remainder of the proof of Abu-Khzam et al.\ carries over without further modification, and we again obtain an instance of {\problemISI}[VC] that is equivalent to the original instance of {\problemPC}.

By inspecting the reduction of Abu-Khzam et al.\ and the above augmentation, we note that the construction of the pattern graph $G_2$ is actually independent of $G$, and only depends on $k$. So denote by $G_2^\ell$ the pattern graph constructed when $k = \ell$. Moreover, we note that the above isomorphism between $\tilde{P}$ and $P$ can only occur when their length is the same. Hence, an induced subgraph of $G_1$ is isomorphic to $G_2^\ell$ only if $\ell = k$.
Now let $\Pi$ be the set of graphs isomorphic to $G_2^\ell$ for any $\ell \geq 1$. 

To complete the argument, consider any algorithm for {\problemPfDV} running in time $2^{o(K \log K)}\mathrm{poly}(n)$. Let $(G,k)$ be any instance of {\problemPC}. Apply the reduction of Abu-Khzam et al.\ to $G$ with the above augmentation to obtain the host graph $G_1$. As argued above, $G_1$ contains an induced subgraph isomorphic to $G_2^\ell$ for any $\ell$ if and only if $G$ has a permutation clique of size $k$. Hence, the instance of {\problemPfDV} for the constructed property $\Pi$ with input graph $G_1$ has solution size at least~$1$ if and only if $G_1$ contains an induced subgraph isomorphic to $G_2^\ell$ for any $\ell$, if and only if $G$ has a permutation clique of size $k$. Hence, using this transformation on an instance of {\problemPC} and then applying the assumed algorithm for {\problemPfDV}, we obtain an algorithm for {\problemPC} with running time $2^{o(k \log k)}\mathrm{poly}(n)$. Such an algorithm cannot exist unless ETH fails.
\end{proof}

Next, we look to further improve the bound of Theorem~\ref{thm:polyFPT}. Note that so far, we have made no use at all of the characterization by few adjacencies of $\Pi$, as in Theorem~\ref{TheoremPiFree}. We now argue that there may be graphs in $\Pi$ that cannot occur in $G$ simply because it would not fit with the vertex cover.

\begin{lemma}
    If $\Pi$ is a graph property such that
    \begin{enumerate}[noitemsep,label={(\roman*)}]
	    \item every graph in $\Pi$ is connected and contains at least one edge, and
	    \item $\Pi$ is characterized by $c_\Pi$ adjacencies,
    \end{enumerate}
    and $G$ is some graph with vertex cover $X$, $|X| = K$, and $S \subseteq V(G)$ some vertex set. Then $G[V(G) \setminus S]$ is $\Pi$-free if and only if $G[V(G) \setminus S]$ is $\Pi'$-free, where $\Pi'\subseteq \Pi$ contains only those graphs in $\Pi$ with $\leq (c_\Pi + 1)K$ vertices.\label{lemma:cpiConnectedBounded}
\end{lemma}
\begin{proof}
        Assume the preconditions in the lemma. When $G[V(G) \setminus S]$ is $\Pi$-free, it must also be $\Pi'$-free, as $\Pi' \subseteq \Pi$ by definition.
	
	    Now assume that $G[V(G) \setminus S]$ is not $\Pi$-free, let us say some $H\in \Pi$ occurs in the graph. Consider a vertex $v$ of the graph $H$. Because $\Pi$ is characterized by $c_\Pi$ adjacencies, there exists a set $D \subseteq V(H)$ with $|D| \leq c_\Pi$ such that changing the adjacencies between $v$ and $V(H)\setminus D$ does not change the presence of $H\in \Pi$. Remove all adjacencies existing between $v$ and $V(H)\setminus D$. Then $\deg(v) \leq c_\Pi$. Our new version of $H$ is still contained in $\Pi$, so we can repeat this process for every vertex in $H$. But then every vertex $v$ in $H$ has $\deg(v) \leq c_\Pi$. Given that $H$ can contain at most $K$ vertices of the vertex cover in $G$, each with degree at most $c_\Pi$, we know that this edited version of $H$ can have at most $(c_\Pi + 1)K$ vertices. Although this version of $H$ is still in $\Pi$, it might not exactly be in $G$, as we might have deleted essential adjacencies. However, changing all the adjacencies did not increase or decrease the number of vertices, as every graph in $\Pi$ is connected and $H$ remains in $\Pi$ at every step. Therefore, every $H$ occurring in $G[V(G) \setminus S]$ is in $\Pi'$, and so the graph is not $\Pi'$-free either.
\end{proof}

The precondition that every graph in $\Pi$ is connected is necessary to obtain this result. If not every graph in $\Pi$ is connected, the removal of adjacencies might leave a vertex without edges, but then this vertex might still be required for the presence in $\Pi$, which is problematic. Seeing that we want to bound the size of possible graphs in $\Pi$, to be able to bound the size of graphs we need that the vertex cover together with $c_\Pi$ gives us information on the size of the graph, which is not the case for a disjoint union of graphs.

We can use Lemma~\ref{lemma:cpiConnectedBounded} in combination with Theorem~\ref{thm:PiFreeP(K)} to obtain a new result. Alternatively, using a streaming version of the algorithm instead of the non-streaming one, immediately also provides a streaming result.

\begin{theorem}
    Given a graph $G$ with vertex cover $X$, $|X| = K$, if $\Pi$ is a graph property such that
	\begin{enumerate}[noitemsep,label={(\roman*)}]
		\item every graph in $\Pi$ is connected and contains at least one edge, and
		\item $\Pi$ is characterized by $c_\Pi$ adjacencies, and
		\item we have explicit knowledge of $\Pi' \subseteq \Pi$, which is the subset of $q$ graphs of at most size $(c_\Pi + 1)K$ that are vertex-minimal with respect to $\Pi$,
    \end{enumerate}
    then \problemPfDV{} can be solved using $\O(q \cdot 2^K \cdot ((c_\Pi + 1)K)^K \cdot K! \cdot K \cdot ((c_\Pi + 1)K)! \cdot((c_\Pi + 1)K)^2\cdot n)$ time. Assuming $c_\Pi \geq 1$ this can be simplified to $\O(q \cdot 2^K \cdot {c_\Pi}^K \cdot K^{K + 3} \cdot K! \cdot (c_\Pi K)! \cdot n)$ time. In the streaming setting, \problemPfDV{} can be solved using $\O(q \cdot 2^K \cdot c_\Pi^K \cdot K^{K + 2} \cdot K! \cdot (c_\Pi K)!)$ passes in the AL streaming model, using $\Ot(q \cdot (c_\Pi \cdot K)^2)$ space.\label{thm:PiFreeFPTandStream}
\end{theorem}
\begin{proof}
        Given some $\Pi$ characterized by $c_\Pi$ adjacencies, where every graph in $\Pi$ is connected, we can see that through Lemma~\ref{lemma:cpiConnectedBounded} we only need to consider those graphs with size $\leq (c_\Pi + 1)K$ in $\Pi$, and with this subset, using Theorem~\ref{thm:PiFreeP(K)} where $p(K) = (c_\Pi + 1)K$, the fpt part of the theorem follows.
        
        For the streaming part, instead of applying Algorithm~\ref{alg:hfreefpt} we apply Algorithm~\ref{alg:hfreestream}. The main factor $q \cdot (c_\Pi \cdot K)^2$ of the memory usage comes from the fact that we need to explicitly store $\Pi'$.
\end{proof}

The required explicit knowledge of $\Pi'$ might give memory problems. That is, we have to store $\Pi'$ somewhere to make this algorithm work, which takes $\Ot(q \cdot (c_\Pi + K)^2)$ space. Note that $q$ can range up to $K^{\O(K)}$. We adapt the streaming algorithm to the case when we have oracle access to~$\Pi$ in Section~\ref{sec:noexplicitpi}.

\subsection{\texorpdfstring{$\Pi$-free Deletion without explicit $\Pi$}{Pi-free Deletion without explicit Pi}}\label{sec:noexplicitpi}

One of the issues in Section~\ref{sec:DirectFPT} is that having the graph property $\Pi$ explicitly saved might cost us a lot of memory. To circumvent this, we can assume to be working with some \emph{oracle}, which we can call to learn something about $\Pi$. In general, if we have an oracle algorithm $\A$, let us assume that it takes a graph $G$ as input as a stream. We then denote $P_\A(n), M_\A(n)$ as respectively the number of passes and the memory use of the oracle algorithm $\A$ when called on a streamed (sub)graph $G$ with $n$ vertices. In this section, we will discuss two different oracles and their use for \problemPfDV{}. We also assume that the graphs in $\Pi$ have a maximum size $\nu$, and that $\nu$ is known.

As the oracle algorithms take a stream as an input, but we usually might want to pass a subgraph of $G$, we require a full pass over the stream for each call to an oracle algorithm. This way, we can select exactly the sub-stream that corresponds to the graph we wish to pass to the oracle algorithm while being memory-less (for each edge we only need to decide whether or not it should appear in the oracle input, and pass it to it if is). Actually, if the oracle algorithm uses multiple passes over its input, we need to generate this stream every time it does so.

The first oracle model is where our oracle algorithm $\A_1$, when called on a graph $G$, returns whether or not $G \in \Pi$. Our approach then has to be different from Algorithm~\ref{alg:hfreestream}, as in Algorithm~\ref{alg:hfreestream} we rely on knowing whether a part of the vertex cover is contained in some $H \in \Pi$. This is not information the oracle can give us. The general idea is the following: we still branch on what part of the vertex cover is in the solution, and consider every subset of (the remaining part of) the vertex cover in each branch. To avoid having to test every combination of vertices outside the vertex cover together with this subset, we consider the notion of \emph{twin vertices}. Two vertices $u,v \notin X$, where $X$ is a vertex cover, are called \emph{twins} when $N(u) = N(v) \subseteq X$, their neighbourhoods are equal. If $EC$ is a set of vertices where each pair $u,v$ are twins, and $EC$ is maximal under this property, we call $EC$ an \emph{equivalence class}. When trying to test which graphs might be in $\Pi$, we can ignore twin vertices if we have tested one of them before. Also, if we delete a vertex from an equivalence class, we may need to delete the entire equivalence class. We can identify twins easily in the \textsc{AL} streaming model. The issue here is that saving the equivalence classes is not memory efficient. Nonetheless, we use this idea here in Algorithm~\ref{alg:pifreeStreamOracle1}.

In Algorithm~\ref{alg:pifreeStreamOracle1}, the set $EC$ saves the sizes of each equivalence class, and can be seen as a set of key-value pairs $(key,val)$, where $key$ is a $K$-bit string representing the adjacencies towards the vertex cover, and $val$ is the number of vertices in this equivalence class. Notice that we can find the set $EC$ using a single pass over the stream. This can be done by, for each vertex, locally saving its adjacencies as a $K$-bit string, and then finding and incrementing the correct counter in $EC$. We only need one pass to find $EC$ in its entirety.

With slight abuse of notation, we write $\textsc{Dict}_{EC_{\leq k}}$ to denote a dictionary ordering on choosing $\leq k$ `vertices' out of the $2^K$ equivalence classes such that if an equivalence class $key$ is chosen twice, then $val$ is at least 2. This is essentially enumerating picking $\leq k$ vertices out of $V \setminus X$ except that we do not pick vertices explicitly, but we pick the equivalence classes they are from. An entry of this ordering is a set of key-value pairs $(key,count)$, such that $key$ corresponds to some equivalence class, and $count$ is the number of vertices we pick out of this equivalence class.

\begin{algorithm}[!ht]
	\caption{\textsc{$\Pi$-free Deletion with $\A_1$}(Graph $G = (V,E)$ in the AL model, integer $\ell$, integer $\nu$, Vertex Cover $X\subseteq V(G)$)}
	\label{alg:pifreeStreamOracle1}
	\begin{algorithmic}[1]
		\State $S \gets \textsc{First}(\X_{\leq \ell})$
		\While{$S \in \X_{\leq \ell}, S \neq \spadesuit$}
		\State $Y \gets X\setminus S$
		\State $S' \gets S$
		\If{$\forall Y' \subseteq Y :$ \Call{$\A_1$}{$Y'$} = false} \Comment{Test if $Y$ is $\Pi$-free}
		\State $EC \gets$ Count the sizes of the equivalence classes of vertices in $V \setminus X$ with their adjacencies towards $Y$ \Comment{Use a pass for this}
		\If{\Call{Search}{$S'$, $Y$, $EC$}} \Return YES \EndIf
		\EndIf
		\State $S \gets \textsc{Dict}_{\X_{\leq \ell}}(\textsc{Next}(S))$
		\EndWhile
		\State \Return NO
		
		\Statex
		
		\Function{Search}{solution set $S$, forbidden set $Y \subseteq X$, equivalence class sizes $EC$}
		\State $J \gets \textsc{First}(\Y_{\leq \nu})$
		\State $I \gets \textsc{First}(EC_{\leq (\nu - |J|)})$
		\While{$J \in \Y_{\leq \nu}, J \neq \spadesuit$}
		\While{$I \in EC_{\leq (\nu - |J|)}, I \neq \spadesuit$}
		
		\If{\Call{$\A_1$}{$I, J$}} \Comment{$I \cup J \in \Pi$}
		\If{$|S| \geq \ell$} \Return NO \Comment{No budget to branch}
		\Else
			\ForEach{$(k,count) \in I$}
			\State Update $EC$ such that $(k,val) \gets (k, count - 1)$ \Comment{Remove all but $count-1$ vertices from class $k$}
			\If{$|S| + val - (count - 1) > \ell$} \Return NO
			\Else \State $VS \gets$ Find $val - (count -1)$ vertices that belong to class $k$ \label{algline:val-count-1}\Comment{Uses a pass}
			\If{\Call{Search}{$S \cup VS$, $Y$, $EC$}} \Return YES \Comment{Branch}\EndIf\EndIf
			\EndFor
		\EndIf
		\EndIf
		\State $I \gets \textsc{Dict}_{EC_{\leq (\nu - |J|)}}(\textsc{Next}(I))$
		\EndWhile
		\State $J \gets \textsc{Dict}_{Y_{\leq \nu}}(\textsc{Next}(J))$
		\State $I \gets \textsc{First}(EC_{\leq (\nu - |J|)})$
		\EndWhile
		\State \Return YES
		\EndFunction
	\end{algorithmic}
\end{algorithm}

\begin{theorem}
    \label{thm:pifreeStreamOracle1}
	If $\Pi$ is a graph property such that the maximum size of graphs in $\Pi$ is $\nu$, and $\A_1$ is an oracle algorithm that, when given subgraph $H$ on $h$ vertices, decides whether or not $H$ is in $\Pi$ using $P_{\A_1}(h)$ passes and $M_{\A_1}(h)$ bits of memory, then Algorithm~\ref{alg:pifreeStreamOracle1} solves \problemPfDV{} on a graph $G$ on $n$ vertices given as an AL stream with a vertex cover of size $K$ using $\O(3^K \nu^{K+1} 2^{\nu K} [1 + P_{\A_1}(\nu)])$ passes and $\O(2^K (K + \log n) + \nu \log n + M_{\A_1}(\nu))$ bits of memory.
\end{theorem}
\begin{proof}
        Let us argue on the correctness of Algorithm~\ref{alg:pifreeStreamOracle1}. In essence, the algorithm tries every option of how an occurrence of a graph in $\Pi$ can occur in $G$, by enumerating all subsets of the vertex cover ($J$) combined with vertices from outside the vertex cover ($I$). This process is optimized by use of the equivalence classes, removing multiple equivalent vertices at once to eliminate such an occurrence. If the algorithm returns YES, then for every combination of $I$ and $J$, no occurrence of a graph in $\Pi$ was found or branching happened on any such occurrence. Each branch eliminates the found occurrence of a graph in $\Pi$, as enough vertices are removed from an equivalence class to make the same occurrence impossible. As removing vertices from the graph cannot create new occurrences of graphs in $\Pi$, this means that every occurrence found was removed, and no new occurrences were created. But then there is no occurrence of a graph in $\Pi$, as every combination of such a graph occurring was tried. So the graph is $\Pi$-free. If the algorithm returns NO, then in every branch at some point we needed to delete vertices to remove an occurrence of a graph in $\Pi$ but had not enough budget left. Removing any less than $val - (count - 1)$ vertices in line~\ref{algline:val-count-1} results in at least $count$ vertices of that equivalence class remaining, which means the same occurrence of a graph in $\Pi$ persists. Therefore, the conclusion that we do not have enough budget is correct, and so there is no solution to the instance. We can conclude that the algorithm works correctly.
	
	    Let us analyse the space usage. Almost all sets used are bounded by $K$ entries. The exceptions are the sets $EC$, $I$, and $J$. $J$ contains at most $\nu$ vertices, and so uses at most $\O(\nu \log n)$ bits of space. $EC$ contains at most $2^K$ key-value pairs, each using $K + \log n$ bits, so $EC$ uses $\O(2^K(K + \log n))$ bits. The set $I$ contains only subsets of $EC$. The oracle algorithm  $\A_1$ uses at most $M_{\A_1}(\nu)$ bits. Therefore, the memory usage of this algorithm is $\O(2^K(K + \log n) + \nu \log n + M_{\A_1}(\nu))$ bits of space. Note that we branch on at most $\nu$ options together spanning at most $\nu$ vertices, which is a constant factor in the memory use (for remembering what we branch on when returning out of recursion). The algorithm also repeats the process for each $I$ and $J$ in each branch to avoid having to keep track of these sets in memory when returning from recursion.
	
	    It remains to show the number of passes used by the algorithm. Let it be clear that the number of passes made by the algorithm is dominated by the number of calls made to $\A_1$, which requires at least one pass. The number of calls made is heavily dependent on the total number of branches in the algorithm, together with how many options the sets $I$ and $J$ can span. This total can be concluded to be $\O(3^K \nu^{K+1} 2^{\nu K})$. Let us elaborate on this. The factor $3^K$ comes from the fact that any vertex in the vertex cover is either in $J$, in $S$, or in $Y$ (and not in $J$). The factor $\nu^K$ comes from the worst case branching process, as we branch on at most $\nu$ different deletions, and we branch at most $\ell \leq K$ times. The remaining factor is the number of options we have for the set $I$, which picks, for each size $i$ between $1$ and $\nu$, $i$ times between at most $2^K$ options (the equivalence classes). This number of options is bounded by $\O(\nu 2^{\nu K})$. This means the total number of passes is bounded by $\O(3^K \nu^{K+1} 2^{\nu K} [1 + P_{\A_1}(\nu)])$.
\end{proof}

Next, we look at a different oracle model, as the memory complexity of $\O(2^K K)$ is not ideal.

The second oracle model is where our oracle algorithm $\A_2$, when called on a graph $G$, returns whether or not $G$ is induced $\Pi$-free (that is, it returns YES when it is induced $\Pi$-free, and NO if there is an occurrence of a graph in $\Pi$ in $G$). The advantage of this model is that we can be less exact in our calls, namely by always passing the entire vertex cover (that is, the part that is not in the solution) together with some set of vertices. The idea for the algorithm with this oracle is the following: we call on the oracle with small to increasingly large sets, such that whenever we get returned that this set is not $\Pi$-free, we can delete a single vertex to make it $\Pi$-free. This can be done by first calling the oracle on the vertex cover itself, and then on the vertex cover together with a single vertex for every vertex outside the vertex cover, then with two, etc.. The problem with this method is the complexity in $n$ that will be present, but as we know the maximum size of graphs in $\Pi$, $\nu$, this is still a bounded polynomial.

We give the algorithm here as Algorithm~\ref{alg:pifreeStreamOracle2}, and we note that it is similar to the \textsc{$\Pi$-free Modification} algorithm of Cai~\cite{CaiModification}.

\begin{algorithm}[!ht]
	\caption{\textsc{$\Pi$-free Deletion with $\A_2$}(Graph $G = (V,E)$ in the streaming model, integer $\ell$, integer $\nu$, Vertex Cover $X\subseteq V(G)$)}
	\label{alg:pifreeStreamOracle2}
	\begin{algorithmic}[1]
		\State $S \gets \textsc{First}(\X_{\leq \ell})$
		\While{$S \in \X_{\leq \ell}, S \neq \spadesuit$}
		\State $Y \gets X\setminus S$
		\State $S' \gets S$
		\If{\Call{$\A_2$}{$Y$}}
		\If{\Call{Search}{$S'$, $Y$, $\emptyset$}} \Return YES \EndIf
		\EndIf
		\State $S \gets \textsc{Dict}_{\X_{\leq \ell}}(\textsc{Next}(S))$
		\EndWhile
		\State \Return NO
		
		\Statex
		
		\Function{Search}{solution set $S$, forbidden set $Y \subseteq X$, set $I$}
			\If{$I = \emptyset$} $I \gets \textsc{First}((V\setminus X)_{\leq \nu})$
			\Else \State $I \gets \textsc{Dict}_{(V\setminus X)_{\leq \nu}}(\textsc{Next}(I))$\EndIf
			\If{$I = \spadesuit$} \Return YES
			\ElsIf{$I \cap S \neq \emptyset$} \Return \Call{Search}{$S$, $Y$, $I$} \Comment{This $I$ is invalid}
			\ElsIf{\Call{$\A_2$}{$Y \cup I$}} \Return \Call{Search}{$S$, $Y$, $I$} \Comment{$Y \cup I$ is $\Pi$-free}
			\ElsIf{$|S| = \ell$} \Return NO \Comment{No budget to branch}
			\Else
			\ForEach{$v \in I$}
				\If{\Call{Search}{$S \cup \{v\}$, $Y$, $I$}} \Return YES \Comment{Branch, this makes $Y \cup I$ $\Pi$-free}\EndIf
			\EndFor
			\EndIf
			\State \Return NO
		\EndFunction
	\end{algorithmic}
\end{algorithm}

We are now ready to prove the complexity of Algorithm~\ref{alg:pifreeStreamOracle2}. Notice that we do not explicitly state what streaming model we use, as this is dependent on the type of stream the oracle algorithm accepts. So if the oracle algorithm works on a streaming model $X$, then so does Algorithm~\ref{alg:pifreeStreamOracle2}.

\begin{theorem}
    \label{thm:pifreeStreamOracle2}
	If $\Pi$ is a graph property such that the maximum size of graphs in $\Pi$ is $\nu$, and $\A_2$ is an oracle algorithm that, when given subgraph $H$ on $h$ vertices, decides whether or not $H$ is $\Pi$-free using $P_{\A_2}(h)$ passes and $M_{\A_2}(h)$ bits of memory, then Algorithm~\ref{alg:pifreeStreamOracle2} solves \problemPfDV{} on a graph $G$ on $n$ vertices given as a stream with a vertex cover of size $K$ using $\O(2^K\nu^{K+1}n^\nu [1 + P_{\A_2}(K + \nu)])$ passes and $\O(\nu K \cdot \log n + M_{\A_2}(K + \nu))$ bits of memory.
\end{theorem}
\begin{proof}
        The correctness of the algorithm can be seen as follows. Whenever the algorithm finds a subset of the graph which is not induced $\Pi$-free, it branches on one of the possible vertex deletions. This always makes this subset $\Pi$-free, as in previous iterations every subset with one vertex less was tried, to which the oracle told us that it was $\Pi$-free (otherwise, we would have branched there already, and this subset could not occur in the current iteration). Therefore, the branching process removes the found occurrences of induced $\Pi$ graphs. The algorithm only returns YES when all subsets of at most $\nu$ vertices have been tried as $I$, and as every graph in $\Pi$ has at most $\nu$ vertices, this includes every possibility of a graph in $\Pi$ appearing in $G$. Therefore, if at this point we have a set $S$ of at most $\ell$ vertices, then $G[V\setminus S]$ must be $\Pi$-free, and so $S$ is a solution to \problemPfDV{}. If the algorithm returns NO, then for every possible subset of the vertex cover contained in $S$, there is not enough budget to remove all occurrences of induced $\Pi$ graphs, and therefore there is no solution to \problemPfDV{}. We can conclude the algorithm works correctly.
	
	    Let us analyse the space usage. The algorithm requires space for the sets $S$, $S'$, $Y$, $X$, $I$, and for the oracle algorithm $\A_2$. The size of each of the sets $S$, $S'$, $Y$, and $X$ is bounded by $K$ entries, so we use $\O(K \log n)$ bits memory for them. The set $I$ contains at most $\nu$ entries, and therefore requires at most $\O(\nu \log n)$ bits of space. The oracle algorithm is called on graphs of at most $K + \nu$ vertices, and so requires $M_{\A_2}(K + \nu)$ bits of memory. To handle branching correctly, we need to save the set $I$ and all vertices we branch on every time we branch. These sets have a maximum size of $\nu$, and the search tree has a maximum depth of $K$, so this takes $\O(\nu K \log n)$ bits of memory. Therefore, the total memory usage comes down to $\O(\nu K \cdot \log n + M_{\A_2}(K + \nu))$ bits. Seeing as $\nu$ is a constant, we could further simplify this, but we choose to be more explicit here.
	
	    As mentioned, we require a pass over the stream to call on the oracle. Therefore, the number of passes on the stream is dependent on the number of calls to the oracle algorithm $\A_2$. In the worst case, we call on $\A_2$ in every branch for every set $I$. The number of branches is bounded by $\O(2^K \nu^K)$, as this encompasses every subset of the vertex cover and every deletion branch is on at most $\nu$ vertices. There are $\sum_{i=1}^{\nu} {n \choose i} = \O(\nu n^\nu)$ different subsets $I$. This means the total number of calls to $\A_2$ is bounded by $\O(2^K \nu^{K+1} n^\nu)$, which means the total number of passes is bounded by $\O(2^K\nu^{K+1}n^\nu [1 + P_{\A_2}(K + \nu)])$.
\end{proof}

Algorithm~\ref{alg:pifreeStreamOracle2} can also be executed using oracle algorithm $\A_1$, but instead of passing the entire vertex cover each time, we would need to do this for every subset of the vertex cover. This would make the $2^K$ factor in the number of passes a $3^K$ factor (a vertex in the vertex cover can either not be in $Y$, or be in $Y$ and not passed, or in $Y$ and passed to $\A_1$). Let us shortly formalize this in a theorem.

\begin{theorem}
    \label{thm:pifreeStreamOracle1v2}
	If $\Pi$ is a graph property such that the maximum size of graphs in $\Pi$ is $\nu$, and $\A_1$ is an oracle algorithm that, when given subgraph $H$ on $h$ vertices, decides whether or not $H$ is in $\Pi$ using $P_{\A_1}(h)$ passes and $M_{\A_1}(h)$ bits of memory, then Algorithm~\ref{alg:pifreeStreamOracle2} can be adapted to solve \problemPfDV{} on a graph $G$ on $n$ vertices given as a stream with a vertex cover of size $K$ using $\O(3^K\nu^{K+1}n^\nu [1 + P_{\A_1}(\nu)])$ passes and $\O(\nu K \cdot \log n + M_{\A_1}(\nu))$ bits of memory.
\end{theorem}
\begin{proof}
        The only difference between oracle algorithm $\A_1$ and $\A_2$ is that $\A_1$ requires an exact occurrence to be input, while for oracle $\A_2$ a larger set can suffice. As Algorithm~\ref{alg:pifreeStreamOracle2} already enumerates all of the possibilities for vertices outside the vertex cover, only refinement is necessary for how the vertex cover itself is passed to the oracle. At each location where a call happens to $\A_2$ in Algorithm~\ref{alg:pifreeStreamOracle2}, we actually want to replace this by an enumeration of calls for each subset of the vertex cover, similar to how Algorithm~\ref{alg:pifreeStreamOracle1} has a combination of sets $I$ and $J$. By the correctness of Algorithm~\ref{alg:pifreeStreamOracle2}, this approach also leads to a correct solution. The number of calls to $\A_1$ does increase in comparison to $\A_2$, however. We can include the increase in complexity in the already existing branching complexity on the vertex cover. In any branch, any vertex in the vertex cover is either in $S$, or in $Y$ and passed to $\A_1$, or in $Y$ and not passed. This means that the number of calls to $\A_1$ is bounded by $\O(3^K\nu^{K+1}n^\nu)$, and therefore the total number of passes is bounded by $\O(3^K\nu^{K+1}n^\nu [1 + P_{\A_1}(\nu)])$. The memory usage of the algorithm is asymptotically the same as that of Algorithm~\ref{alg:pifreeStreamOracle2}, with $\A_2$ replaced by $\A_1$.
\end{proof}

Another approach in which we do not demand $\nu$ is known comes from the algorithm that Theorem~\ref{thm:generic-upper-fpt} proposes. We work with oracle model $\A_2$, which can check whether a given stream is $\Pi$-free. We can enumerate the $(2^K + 1)\cdot K$ possible solutions by saving $\O(K^2)$ bits corresponding to the deletion of $\O(K)$ vertices of equivalence classes, which each require $\O(K)$ bits to represent. We can enumerate all these $(2^K + 1)\cdot K$ options by using Dictionary Orderings. This immediately provides an algorithm, as for each possible solution, for each pass the oracle requires, we make a pass over the AL stream and pass on to $\A_2$ only the vertices from an equivalence class when the first $i$ have been skipped (where $i$ corresponds to the number of deletions in this equivalence class according to the current proposed solution). The correctness of this process follows from Theorem~\ref{thm:generic-upper-fpt}, and the number of passes comes down to $\O(2^{O(K^2)} \cdot P_{\A_2}(n))$, with a memory usage of $\O(K^2 + M_{\A_2}(n))$ bits.

We have now seen some algorithms for using an oracle to obtain information about $\Pi$ instead of explicitly saving it. This way, we have gotten closer to a memory-optimal algorithm, but the number of passes over the stream has increased significantly in comparison to algorithms from Section~\ref{sec:DirectFPT}.

%% file: ms_oct.tex
\subsection{Odd Cycle Transversal}\label{sec:oct}

Specific forms of \problemPfDV{} allow for improvement over Theorem~\ref{thm:PiFreeFPTandStream}, which we illustrate for the problem of \textsc{Odd Cycle Transversal~[VC]}. Note that odd cycle-free and induced odd cycle-free are equivalent.

\begin{problemParam}
	\problemtitle{\textsc{Odd Cycle Transversal~[VC]}}
	\probleminput{A graph $G$ with a vertex cover $X$, and an integer $\ell$.}
	\problemparameter{The size $K \coloneqq |X|$ of the vertex cover.}
	\problemquestion{Is there a set $S\subseteq V(G)$ of size at most $\ell$ such that $G[V(G)\setminus S]$ contains no induced odd cycles?}
\end{problemParam}

The interest in this problem comes from the FPT algorithm using iterative compression provided in \cite[Section 4.4]{ParameterizedAlgorithmsBook}, based on work by Reed et al.~\cite{OCT_source}. Although Chitnis and Cormode~\cite{ChitnisTheory} have shown how iterative compression can be used in the streaming setting, adapting the algorithm out of Reed et al. seems difficult. The main cause for this is the use of a maximum-flow algorithm, which does not seem to lend itself well to the streaming setting because of its memory requirements. 
Instead, we present the following approach.

It is well known that a graph without odd cycles is a bipartite graph (and thus 2-colourable) and vice versa. 
In the algorithm, we guess what part of the vertex cover is in the solution, and then we guess the colouring of the remaining part. Then vertices outside the vertex cover for which not all neighbours have the same colour must be deleted. This step can be done in one pass if we use the AL streaming model. In the same pass, we can also check if the colouring is valid within the vertex cover. If the number of deletions does not exceed the solution size and the colouring is valid within the vertex cover, then the resulting graph is bipartite and thus odd cycle free.

The total number of guesses comes down to $\O(3^K)$ options, as any vertex in the vertex cover is either in the solution, coloured with colour 1 or coloured with colour 2. This directly corresponds to the number of passes, as only one pass is needed per guessed colouring.

The full algorithm is given in Algorithm~\ref{alg:OCTstream}.
\begin{algorithm}[!htp]
	\caption{\textsc{OCT}(Graph $G = (V,E)$ in the AL model, integer $\ell$, Vertex Cover $X\subseteq V(G)$)}
	\label{alg:OCTstream}
	\begin{algorithmic}[1]
		\State $S \gets \textsc{First}(\X_{\leq \ell})$
		\While{$S \in \X_{\leq \ell}, S \neq \spadesuit$}
			\State $Y \gets X\setminus S$
			\State $Y_1 \gets \textsc{First}(\Y)$
			\State $Y_2 \gets Y \setminus Y_1$
			\While{$Y_1 \subseteq Y, Y_1 \neq \spadesuit$}
				\State $success \gets true, S' \gets S$ \Comment{Reset local values}
				\ForEach{$v\in V \setminus S'$} \Comment{Use one pass}
					\If{$v \in Y$ and $v \in Y_i$} \Comment{$v\in X$}
						\State Check that all neighbours of $v$ in $Y$ are in $Y_{3-i}$\label{algline:OCTVCcheck}
						\State If one is not, $success \gets false$
					\Else \Comment{$v \notin X$}
						\State Check if all neighbours of $v$ in $Y$ are in the same $Y_i$\label{algline:OCTothercheck}
						\State If not, and $|S'| < \ell$, $S' \gets S' \cup \{v\}$
						\State Else, $success \gets false$
					\EndIf
				\EndFor
				\If{$|S'| \leq \ell$ and $success$} \Return YES \EndIf
				\State $Y_1 \gets \textsc{Dict}_{\Y}(\textsc{Next}(Y_1))$ \Comment{Try the next colouring}
				\State $Y_2 \gets Y \setminus Y_1$
			\EndWhile
			\State $S \gets \textsc{Dict}_{\X_{\leq \ell}}(\textsc{Next}(S))$
		\EndWhile
		\State \Return NO
	\end{algorithmic}
\end{algorithm}

\begin{theorem}
    Given a graph $G$ given as an AL stream with vertex cover $X$, $|X| = K$, we can solve \textsc{Odd Cycle Transversal~[VC]} using $\O(3^K)$ passes and $\O(K \log n)$ space.\label{thm:octstream1}
\end{theorem}
\begin{proof}
        We claim that Algorithm~\ref{alg:OCTstream} does exactly this.
        
        Let us first prove the correctness of Algorithm~\ref{alg:OCTstream}. Let $G = (V,E)$ be a graph with vertex cover $X$, $|X| = K$. Let $O$, $|O| \leq \ell$ be a solution for \textsc{Odd Cycle Transversal [VC]}. Denote $Y' = X \cap O$ as the part of $O$ that is contained in the vertex cover. Because Algorithm~\ref{alg:OCTstream} enumerates all possibilities for the set $Y$, at some point it must consider $Y = Y'$. As $O$ is a solution, $G[X \setminus Y']$ must be bipartite, and so admits a proper 2-colouring $Y_1'$, $Y_2'$. For $Y = Y'$, Algorithm~\ref{alg:OCTstream} considers at some point $Y_1 = Y_1'$ and $Y_2 = Y_2'$ ($Y_1' \cup Y_2' = Y' = Y$ because $Y_1'$ and $Y_2'$ form a proper 2 colouring), because it considers all possibilities for the set $Y_1 \subseteq Y$. As $Y_1$ and $Y_2$ now form a proper 2-colouring of $Y$, the check in line~\ref{algline:OCTVCcheck} never fails. The check in line~\ref{algline:OCTothercheck} only fails if a vertex outside of $X$ is adjacent to two different coloured vertices in $Y$. But then this vertex must also be in $O$, as $Y_1$ and $Y_2$ mimic exactly the 2-colouring in $G[X \setminus Y']$. Therefore, Algorithm~\ref{alg:OCTstream} can at least find $O$ as well, which means it returns YES. For the reverse implication, when Algorithm~\ref{alg:OCTstream} returns YES, it has found a set $S$ such that $|S| \leq \ell$ and $G[V \setminus S]$ admits a proper 2-colouring given by deterministically adding vertices outside the vertex cover to $Y_1$ and $Y_2$. But then $S$ is a solution to \textsc{Odd Cycle Transversal [VC]}. We can conclude that Algorithm~\ref{alg:OCTstream} works correctly.
	
    	Let us analyse the memory usage of Algorithm~\ref{alg:OCTstream}. All sets used in the algorithm have size at most $K$ ($\ell \leq K$). The only worry is whether or not the checks in lines~\ref{algline:OCTVCcheck} and \ref{algline:OCTothercheck} require more memory. The first check only requires us to remember in what set the vertex $v$ is contained in, and whether or not we have seen a `wrong' colour yet, which should only take a constant number of bits. The second check merely needs to remember in what set all neighbours up until this point were, which should also only take a constant number of bits. Therefore, the algorithm uses $\O(K \log n)$ bits of memory.
	
    	Let us analyse the number of passes of Algorithm~\ref{alg:OCTstream} does over the stream. Firstly, the for-loop over all vertices only requires a single pass because the checks only need to know what all the neighbours of the current vertex are, which is what the AL stream gives us. The total number of times this for-loop can be executed is bounded by $\O(3^K)$, as any vertex in the vertex cover can either be in $S$, in $Y_1$ or in $Y_2$. We can conclude that Algorithm~\ref{alg:OCTstream} uses $\O(3^K)$ passes over the graph stream.
\end{proof}

If we think about this algorithm, we can notice that often the colouring we guess on the vertex cover is invalid. 
An alternative approach follows by noting that a connected component within the vertex cover can only have two possible valid colourings. We can exploit this to decrease the number of passes when the number of connected components in the vertex cover is low. This comes at a price: to easily find components of the vertex cover, we store it in memory, which increases the memory complexity. Alternatively, we can use $\O(K)$ passes to find the connected components of the vertex cover in every branch. We formalize this in Algorithm~\ref{alg:OCTstreamCC} and Theorem~\ref{thm:OCTstreamCC}.

\begin{algorithm}[!htp]
	\caption{\textsc{OCT-CC}(Graph $G = (V,E)$ in the AL model, integer $\ell$, Vertex Cover $X\subseteq V(G)$)}
	\label{alg:OCTstreamCC}
	\begin{algorithmic}[1]
		\State $S \gets \textsc{First}(\X_{\leq \ell})$
		\While{$S \in \X_{\leq \ell}, S \neq \spadesuit$}
			\State $Y \gets X\setminus S$
			\State Use a pass to find and save the edges in $Y$
			\State Find the connected components and their two 2-colourings
			\State If this fails, move to the next option for $S$
				\ForEach{Combination of colourings of the CCs} \Comment{Does $2^{\#\text{CCs}}$ iterations}
					\State $success \gets true, S'\gets S$ \Comment{Reset local values}
					\ForEach{$v\in V \setminus (X \cup S')$} \Comment{Use one pass}
						\State Check if all neighbours of $v$ in $Y$ have the same colour
						\State If not, and $|S'| < \ell$, $S' \gets S' \cup \{v\}$
						\State Else, $success \gets false$
					\EndFor
					\If{$|S'| \leq \ell$ and $success$} \Return YES \EndIf
				\EndFor
			\State $S \gets \textsc{Dict}_{\X_{\leq \ell}}(\textsc{Next}(S))$
		\EndWhile
		\State \Return NO
	\end{algorithmic}
\end{algorithm}

\begin{theorem}\label{thm:OCTstreamCC}
    Given a graph $G$ given as an AL stream with vertex cover $X$, $|X| = K$, Algorithm~\ref{alg:OCTstreamCC} solves \textsc{Odd Cycle Transversal [VC]} using $\O(3^K)$ passes and $\O(K^2 \log n)$ bits of memory.
\end{theorem}
\begin{proof}
        The correctness of Algorithm~\ref{alg:OCTstreamCC} quickly follows from the correctness of Algorithm~\ref{alg:OCTstream}. Where Algorithm~\ref{alg:OCTstream} enumerates all 2-colourings of $Y$, Algorithm~\ref{alg:OCTstreamCC} only enumerates those which are feasible 2-colourings for $Y$ (combinations of 2-coloured components). This means Algorithm~\ref{alg:OCTstreamCC} only leaves out colourings which are not feasible anyway, which means that it still works correctly.
	
	    The number of passes is heavily dependent on the amount of connected components in $G[Y]$ in an iteration. But, in worst case, this is $|Y|$, which would mean it considers all 2-colourings of $Y$. But this is a worst case where the behaviour exactly mimics that of Algorithm~\ref{alg:OCTstream}, and so the worst case number of passes is the same.
	
	    Let it be clear that the memory use is $\O(K^2 \log n)$ bits, as we save (a part of) the vertex cover with edges to enable easy colouring and component finding. The rest of the sets in the algorithm use $\O(K \log n)$ bits of memory.
\end{proof}

Let us denote again that Algorithm~\ref{alg:OCTstreamCC} on paper is strictly worse than Algorithm~\ref{alg:OCTstream} in worst case complexity. However, we believe there to be many cases where Algorithm~\ref{alg:OCTstreamCC} can outperform Algorithm~\ref{alg:OCTstream} because of its behaviour of clever enumeration instead of trying all possibilities. We also want to mention that, instead of saving the vertex cover using $\O(K^2 \log n)$ bits of memory, we could find the connected components of the vertex cover in $\O(K)$ passes in each branch. This increases the number of passes by a factor $K$, but remains memory-optimal, using $\O(K \log n)$ bits of memory.

%% file: ms_lowerbounds.tex
\section{Lower Bounds}\label{sec:lowerbounds}

We show lower bounds for \textsc{$\Pi$-free Deletion}. 
To prove lower bounds for streaming, we can show reductions from problems in communication complexity, as first shown by Henzinger et al.~\cite{HenzingerStreams}. An example of such a problem is \textsc{Disjointness}.

\begin{problemP}
	\problemtitle{\textsc{Disjointness}}
	\probleminput{Alice has a string $x \in \{0,1\}^n$ given by $x_1x_2\ldots x_n$. Bob has a string $y \in \{0,1\}^n$ given by $y_1y_2\ldots y_n$.}
	\problemquestion{Bob wants to check if $\exists 1\leq i \leq n$ such that $x_i = y_i = 1$. (Formally, the answer is NO if this is the case.)}
\end{problemP}

The following proposition is given and used by Bishnu et al.~\cite{SaketVCNr}, and gives us one important consequence of reductions from a problem in communication complexity to a problem for streaming algorithms.

\begin{proposition}{(Rephrasing of item $(ii)$ of \cite[Proposition~5.6]{SaketVCNr})}\label{prop:DisjReduction}
	If we can show a reduction from \textsc{Disjointness} to problem $\Pi$ in streaming model $\M$ such that the reduction uses a 1-pass streaming algorithm of $\Pi$ as a subroutine, then any streaming algorithm working in the model $\M$ for $\Pi$ that uses $p$ passes requires $\Omega(n/p)$ bits of memory, for any $p \in \N$ \cite{ChitnisAnnouncement,BishnuFundamentalGeometric,AgarwalSpatialScan}.
\end{proposition}

The structure of these reductions is relatively simple: have Alice and Bob construct the input for a streaming algorithm depending on their input to \textsc{Disjointness}. If we do this in such a manner that the solution the streaming algorithm outputs gives us exactly the answer to \textsc{Disjointness}, we can conclude that the streaming algorithm must abide the lower bound of \textsc{Disjointness}.

Chitnis et al.~\cite[Theorem~6.3]{ChitnisEsfandiariSampling} prove hardness for many $\Pi$, those that obide to a small precondition. However, Chitnis et al. do not describe in their reduction how Alice and Bob give their `input' as a stream to the algorithm for \textsc{$\Pi$-free Deletion}, and thus it would apply only to the EA streaming model. However, if we observe the proof closely, we can see it extends to the \textsc{VA} model.

We would also like it to extend to the \textsc{AL} model. However, this requires a slightly stronger precondition on the graph class $\Pi$.

\begin{theorem}\label{thm:PiFreeDeletionALLowerBound}
	If $\Pi$ is a set of graphs such that each graph in $\Pi$ is connected, and there is a graph $H \in \Pi$ such that
	\begin{itemize}[noitemsep]
		\item $H$ is a minimal element of $\Pi$ under the operation of taking subgraphs, i.e., no proper subgraph of $H$ is in $\Pi$, and
		\item $H$ has at least two \textbf{disjoint} edges,
	\end{itemize}
	then any $p$-pass (randomized) streaming algorithm working on the \textsc{AL} streaming model for \textsc{$\Pi$-free Deletion~[$\ell$]} needs $\Omega(n/p)$ bits of space.
\end{theorem}
\begin{proof}
		We add onto the proof of \cite[Theorem~6.3]{ChitnisEsfandiariSampling}, by specifying how Alice and Bob provide the input to the $p$-pass streaming algorithm.
		
		Let $H$ be a minimal graph in $\Pi$ which has at least two disjoint edges, say $e_1$ and $e_2$. Let $H' \coloneqq H\setminus\{e_1, e_2\}$. Create as an input for the streaming algorithm $n$ copies of $H'$, where in copy $i$ we add the edges $e_1$ and $e_2$ if and only if the input of \textsc{Disjointness} has a $1$ for index $i$ for Alice and Bob respectively.
		
		As $e_1$ and $e_2$ are disjoint, $e_2$ is incident on two vertices $v,w$ which are not incident to $e_1$. For every pass the algorithm requires, we do the following. We provide all the copies of $H$ as input to the streaming algorithm by letting Alice input all vertices $V(H) \setminus \{v,w\}$ as an AL stream. Note that Alice has enough information to do this, as the vertices incident on the edge $e_2$ in each copy of $H$ is never included in this part of the stream. Then Alice passes the memory of the streaming algorithm to Bob, who inputs the edges incident to the vertices $v,w$ for each copy of $H$ (which includes $e_2$ if and only if the respective bit in the input of \textsc{Disjointness} is 1). This ends a pass of the stream.
		
		Note that Alice and Bob have input the exact specification of a graph as described by Chitnis et al.~\cite[Theorem~6.3]{ChitnisEsfandiariSampling}, but now as a AL stream. Hence, the correctness follows.\qed
\end{proof} 

Theorem~\ref{thm:PiFreeDeletionALLowerBound} provides a lower bound for, for example, \textsc{Even Cycle Transversal~[$\ell$]} (where $\Pi$ is the set of all graphs that contain a $C_4, C_6, \ldots$), and similarly \textsc{Odd Cycle Transversal~[$\ell$]} and \textsc{Feedback Vertex Set~[$\ell$]}. Theorem~\ref{thm:PiFreeDeletionALLowerBound} does not hold for the scenario where $\Pi$ contains only stars.

\begin{figure}
	\centering
	\includegraphics[scale=.5]{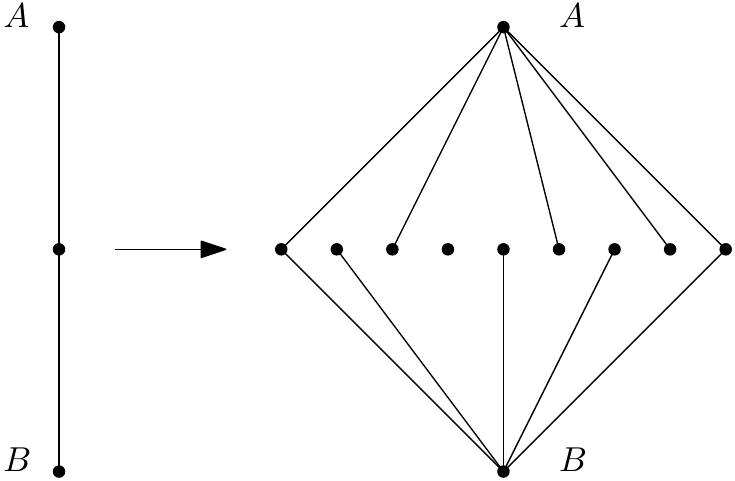}
	\caption{A reduction technique which we call a \emph{double fan}. The right construction can imitate the input to the \textsc{Disjointness} instance, forming the left construction if and only if the answer to \textsc{Disjointness} is NO.}
	\label{fig:doublefan}
\end{figure}

Notice that the lower bound proof makes a construction with a vertex cover size linear in $n$. Therefore, these bounds do not hold when the vertex cover size is bounded. We can prove lower bounds with constant vertex cover size for \textsc{$H$-free Deletion} with specific requirements on $H$, for the EA and VA models. These results follow by adapting the known lower bound construction by Bishnu et al.~\cite{SaketVCNr}. The idea is as follows. We take a copy of $H$, and extend some path of three vertices to the construction in Figure~\ref{fig:doublefan}, which we will call a \emph{double fan}. The $n$ vertices in the overlap of the two fans we will call the \emph{center vertices}. The idea of this construction is that both Alice and Bob input the edges of one of the fans in the double fan, as those edges will be determined by the input to \textsc{Disjointness}. If there is some $1 \leq i \leq n$ such that the $i$-th bit in both Alice's and Bobs input is 1, then the double fan will have a completed path from $A$ to $B$, creating an induced copy of $H$ in the graph. If we make sure the budget is $\ell=0$, then the answer to \textsc{$H$-free Deletion} must be NO if and only if the answer to \textsc{Disjointness} is NO. If we manage this, Proposition~\ref{prop:DisjReduction} gives us a lower bound result.

It is important to note that this construction does not always make a correct reduction. Most importantly, it innately has trouble if $H$ is some star. This is because the double fan construction can form exactly $H$ even if Bobs input exists entirely of zeroes. Therefore, we must be careful to form conditions that exclude stars from the lower bounds proofs. Also, we cannot work on the \textsc{AL} model in this construction. That would mean that for each center vertex, we require some information from both Alice and Bob to construct the stream, which is something that neither Alice nor Bob can do. Avoiding this by making the center vertices edges increases the vertex cover size linearly.

\begin{theorem}
    \label{thm:HfreeLowerBoundVC}
	If $H$ is a connected graph with at least 3 edges and a vertex of degree 2, then any algorithm for solving \textsc{$H$-free Deletion [VC]} on a graph $G$ with $K \geq |\textsc{VC}(H)| + 1$ requires  $\Omega(n/p)$ bits when using $p$ passes in the VA/EA models, even when the solution size $\ell=0$.
\end{theorem}
\begin{proof}
        Similar to the reductions given by Bishnu et al.~\cite{SaketVCNr}, we give a reduction from \textsc{Disjointness} to \textsc{$H$-free Deletion [VC]} in the VA model when the solution size parameter $k=0$. The idea is to build a graph $G$ with bounded vertex cover size, and construct edges according to the input of \textsc{Disjointness}, such that $G$ is $H$-free if and only if the output of \textsc{Disjointness} is YES.
	
	    Let $\A$ be a one-pass streaming algorithm that solves \textsc{$H$-free Deletion [$K$]} in the VA model, such that $|\textsc{VC}(G)| \leq K \leq |\textsc{VC}(H)| + 1$, and the space used is $o(n)$. Let $G$ be a graph with $n + |V(H)| - 1$ vertices consisting of $H$ where a degree-2 vertex in $H$ is expanded to a double fan (i.e., the two adjacencies of this degree-2 vertex correspond to $A$ and $B$, and the degree-2 vertex is replaced by the $n$ center vertices of the double fan). Let $x,y$ be the input strings, consisting of $n$ bits each, of Alice and Bob for \textsc{Disjointness}, respectively.
	
    	Alice exposes to $\A$ all the vertices of $G$, except for the vertex $B$. Here, Alice exposes an edge between $A$ and the $i$-th center vertex of the double fan if and only if the $i$-th bit of $x$ is $1$. Notice that Alice can expose all these vertices according to the VA model, as only the addition of the vertex $B$ will require information of the input of Bob, $y$. If Alice has exposed all vertices of $G$ except for $B$, then she passes the memory of $\A$ to Bob. Bob then exposes the vertex $B$, including an edge between $B$ and the $i$-th center vertex of the double fan if and only if the $i$-th bit of $y$ is 1. This completes the input to $\A$.
	
    	From the construction, observe that $|VC(G)| \leq K \leq |VC(H)| + 1$, as we may need to include both $A$ and $B$ in the vertex cover in $G$ while it was optimal to include the degree-2 vertex in the vertex cover in $H$.
	
    	If the answer to \textsc{Disjointness} is NO, that is, there exists an index $i$ such that $x_i = y_i = 1$, then in $G$ the edges from $A$ and $B$ connect in the $i$-th center vertex, creating an induced copy of $H$ in $G$, and so the graph is not $H$-free. Because $k=0$, \textsc{$H$-free Deletion [$K$]} must also be answered with NO.
    	
    	If the answer to \textsc{Disjointness} is YES, that is, there is no index $i$ such that $x_i = y_i = 1$, then there is no path from $A$ to $B$ through a center vertex in $G$. We will show that there is no induced occurrence of $H$ in $G$. If the degree-2 vertex that we split into the double fan is contained in a cycle in $H$, then now this cycle is no longer present in the graph. As the rest of the graph simply consists of a (partial) copy of $H$, this means there cannot be enough cycles in the graph to get exactly $H$, and so $H$ does not appear in $G$. Otherwise, the degree-2 vertex is not contained in a cycle, and so there is no path between $A$ and $B$. As $H$ has at least three edges, there must be some edge incident to $A$ or $B$ that is not incident to a center vertex. Consider in $H$ the longest path that this degree-2 vertex is contained in. This path must have at least three edges and contain both $A$ and $B$, as $H$ is connected. However, because there is no path between $A$ and $B$, the longest path in $G$ containing $A$ or $B$ must be smaller that the longest path in $H$ containing $A$ and $B$. Hence, the only way for $H$ to occur in $G$ is for this path through $A$ and $B$ to occur somewhere else in $G$. However, this would mean there is now some other path of at least the same length that needs `another place', as it were, to make the induced copy of $H$ appear in $G$. We can see that repeating this process always yields in another path of at least the same length which needs to occur in $G$. However, since we destroyed at least one path of at least this length, all these paths cannot appear in $G$. Hence, the answer to \textsc{$H$-free Deletion [VC]} is YES.
	
    	Now, from Proposition~\ref{prop:DisjReduction} it follows that any algorithm for solving \textsc{$H$-free Deletion [VC]} on a graph $G$ with $K \leq |\textsc{VC}(H)| + 1$ requires at least $n/p$ bits when using $p$ passes in the VA/EA models, even when $\ell=0$. This can be generalized for every $\ell$ by adding $\ell$ disjoint copies of $H$ to $G$, which also increases the vertex cover of $G$ by a constant amount for each copy.\qed
\end{proof}

There are other conditions for which this reduction works as well.

\begin{theorem}
    \label{thm:HfreeLowerBoundVC2}
	If $H$ is a graph with no vertex of degree 1, then any algorithm for solving \textsc{$H$-free Deletion [VC]} on a graph $G$ with $K \geq |V(H)|$ requires $\Omega(n/p)$ bits when using $p$ passes in the VA/EA models, even when the solution size $\ell=0$.
\end{theorem}
\begin{proof}
        Similar to the reductions given by Bishnu et al.~\cite{SaketVCNr}, we give a reduction from \textsc{Disjointness} to \textsc{$H$-free Deletion [VC]} in the VA model when the solution size parameter $k=0$. The idea is to build a graph $G$ with bounded vertex cover size, and construct edges according to the input of \textsc{Disjointness}, such that $G$ is $H$-free if and only if the output of \textsc{Disjointness} is YES.
	
    	Let $\A$ be a one-pass streaming algorithm that solves \textsc{$H$-free Deletion [VC]} in the VA model, such that $|\textsc{VC}(G)| \leq K \leq |V(H)|$, and the space used is $o(n)$. Let $G$ be a graph with $n + |V(H)| - 1$ vertices consisting of $H$ where a vertex of minimal degree in $H$ is expanded to a double fan, i.e., two adjacencies of this vertex correspond to $A$ and $B$, and the vertex is replaced by the $n$ center vertices of the double fan. All other adjacencies of this vertex are connected to all of the center vertices. Note that this is possible, as the vertex has degree at least 2. Let $x,y$ be the input strings, consisting of $n$ bits each, of Alice and Bob for \textsc{Disjointness}, respectively.
	
    	Alice exposes to $\A$ all the vertices of $G$, except for the vertex $B$. Here, Alice exposes an edge between $A$ and the $i$-th center vertex of the double fan if and only if the $i$-th bit of $x$ is $1$. Notice that Alice can expose all these vertices according to the VA model, as only the addition of the vertex $B$ will require information of the input of Bob, $y$. If Alice has exposed all vertices of $G$ except for $B$, then she passes the memory of $\A$ to Bob. Bob then exposes the vertex $B$, including an edge between $B$ and the $i$-th center vertex of the double fan if and only if the $i$-th bit of $y$ is 1. This completes the input to $\A$.
	
    	From the construction, observe that $|VC(G)| \leq K \leq |V(H)|$, as the vertex cover of $G$ can always be bounded by taking all vertices originally in $H$, which covers the edges towards the $n$ center vertices.
	
    	If the answer to \textsc{Disjointness} is NO, that is, there exists an index $i$ such that $x_i = y_i = 1$, then in $G$ the edges from $A$ and $B$ connect in the $i$-th center vertex, creating an induced copy of $H$ in $G$, and so the graph is not $H$-free. Because $\ell=0$, \textsc{$H$-free Deletion [VC]} must also be answered with NO.
	
    	If the answer to \textsc{Disjointness} is YES, that is, there is no index $i$ such that $x_i = y_i = 1$, then there is no path from $A$ to $B$ directly through a center vertex in $G$. We will show that there is no induced occurrence of $H$ in $G$. Name the minimal degree of vertices in $H$ as $d$. Then the center vertices must have degree at most $d-1$, as each center vertex cannot be adjacent to both $A$ and $B$. But as $d$ was the minimal degree in $H$, none of these center vertices can be used for an induced copy of $H$ in $G$. But then $G$ only has $|V(H)| - 1$ vertices remaining to form a copy of $H$, which is impossible. Hence,  the answer to \textsc{$H$-free Deletion [$K$]} is YES.
	
    	Now, from Proposition~\ref{prop:DisjReduction} it follows that any algorithm for solving \textsc{$H$-free Deletion [$K$]} on a graph $G$ with $K \geq |V(H)|$ requires at least $n/p$ bits when using $p$ passes in the VA/EA models, even when $\ell=0$. This can be generalized for every $\ell$ by adding $\ell$ disjoint copies of $H$ to $G$, which also increases the vertex cover of $G$ by a constant amount for each copy.\qed
\end{proof}

The following theorem is a generalization of the above Theorem~\ref{thm:HfreeLowerBoundVC2}.

\begin{theorem}
    \label{thm:HfreeLowerBoundVC3}
	If $H$ is a graph with a vertex of degree at least 2 for which every neighbour has an equal or larger degree, then any algorithm for solving \textsc{$H$-free Deletion [VC]} on a graph $G$ with $K \geq |V(H)|$ requires $\Omega(n/p)$ bits when using $p$ passes in the VA/EA models, even when the solution size $\ell=0$.
\end{theorem}
\begin{proof}
        Similar to the reductions given by Bishnu et al.~\cite{SaketVCNr}, we give a reduction from \textsc{Disjointness} to \textsc{$H$-free Deletion [VC]} in the VA model when the solution size parameter $\ell=0$. The idea is to build a graph $G$ with bounded vertex cover size, and construct edges according to the input of \textsc{Disjointness}, such that $G$ is $H$-free if and only if the output of \textsc{Disjointness} is YES.
	
    	Let $\A$ be a one pass streaming algorithm that solves \textsc{$H$-free Deletion [VC]} in the VA model, such that $|\textsc{VC}(G)| \leq K \leq |V(H)|$, and the space used is $o(n)$. Let $G$ be a graph with $n + |V(H)| - 1$ vertices consisting of $H$ where a vertex degree at least 2 for which every neighbour has an equal or larger degree in $H$ is expanded to a double fan, i.e., two adjacencies of this vertex correspond to $A$ and $B$, and the vertex is replaced by the $n$ center vertices of the double fan. All other adjacencies of this vertex are connected to all of the center vertices. Note that this is possible, as the vertex has degree at least 2. Let $x,y$ be the input strings, consisting of $n$ bits each, of Alice and Bob for \textsc{Disjointness}, respectively.
	
    	Alice exposes to $\A$ all the vertices of $G$, except for the vertex $B$. Here, Alice exposes an edge between $A$ and the $i$-th center vertex of the double fan if and only if the $i$-th bit of $x$ is $1$. Notice that Alice can expose all these vertices according to the VA model, as only the addition of the vertex $B$ will require information of the input of Bob, $y$. If Alice has exposed all vertices of $G$ except for $B$, then she passes the memory of $\A$ to Bob. Bob then exposes the vertex $B$, including an edge between $B$ and the $i$-th center vertex of the double fan if and only if the $i$-th bit of $y$ is 1. This completes the input to $\A$.
	
    	From the construction, observe that $|VC(G)| \leq K \leq |V(H)|$, as the vertex cover of $G$ can always be bounded by taking all vertices originally in $H$, which covers the edges towards the $n$ center vertices.
	
    	If the answer to \textsc{Disjointness} is NO, that is, there exists an index $i$ such that $x_i = y_i = 1$, then in $G$ the edges from $A$ and $B$ connect in the $i$-th center vertex, creating an induced copy of $H$ in $G$, and so the graph is not $H$-free. Because $\ell=0$, \textsc{$H$-free Deletion [VC]} must also be answered with NO.
	
    	If the answer to \textsc{Disjointness} is YES, that is, there is no index $i$ such that $x_i = y_i = 1$, then there is no path from $A$ to $B$ directly through a center vertex in $G$. We will show that there is no induced occurrence of $H$ in $G$. Let us call the vertex that was expanded into the center vertices $v$, and say it has degree $d \geq 2$. Then all the neighbours of this vertex must also have degree at least $d$ in $H$. However, the center vertices in $G$ have degree at most $d-1$, as no center vertex can be adjacent to both $A$ and $B$. Hence, no center vertex can be used for $v$ or any of its neighbours in an induced copy of $H$ in $G$. Consider in $H$ all vertices of degree at least $d$ where the neighbours also have degree at least $d$, and say there are $c$ many of these vertices. In any induced copy of $H$ in $G$, these vertices must still have this relation of degrees. However, as none of the center vertices has degree at least $d$, $G$ contains at most $c-1$ such vertices, which means an induced copy of $H$ cannot occur in $G$. Hence,  the answer to \textsc{$H$-free Deletion [VC]} is YES.
	
    	Now, from Proposition~\ref{prop:DisjReduction} it follows that any algorithm for solving \textsc{$H$-free Deletion [VC]} on a graph $G$ with $K \geq |V(H)|$ requires at least $n/p$ bits when using $p$ passes in the VA/EA models, even when $\ell=0$. This can be generalized for every $\ell$ by adding $\ell$ disjoint copies of $H$ to $G$, which also increases the vertex cover of $G$ by a constant amount for each copy.\qed
\end{proof}

In Theorems~\ref{thm:HfreeLowerBoundVC2} and \ref{thm:HfreeLowerBoundVC3} we only demand that the vertex cover size is at least $|V(H)|$, the number of vertices in $H$. One can wonder if this bound can be tightened, as in Theorem~\ref{thm:HfreeLowerBoundVC}, where we only demand that the vertex cover size is at least $|\textsc{VC}(H)| + 1$. The problem in Theorems~\ref{thm:HfreeLowerBoundVC2} and \ref{thm:HfreeLowerBoundVC3}, is that we might split a vertex of high degree. To get a valid vertex cover without it having linear size in $n$, the only option is to include at least all adjacencies of the center vertices. This makes it that the vertex cover can get a size up to $|V(H)|$, and so this is the only safe demand.

For clarity, we summarize these lower bounds for bounded vertex cover size in a corollary.

\begin{corollary}\label{thm:LowerboundsVCsummary}
    If $H$ is such that either:
    \begin{enumerate}
        \item $H$ is a connected graph with at least 3 edges and a vertex of degree 2, or,
        \item $H$ is a graph with a vertex of degree at least 2 for which every neighbour has an equal or larger degree,
    \end{enumerate}
    then any algorithm for solving \textsc{$H$-free Deletion~[VC]} on a graph $G$ with $K \geq |V(H)|$ requires $\Omega(n/p)$ bits when using $p$ passes in the VA/EA models, even when the solution size $\ell=0$.
\end{corollary}

Corollary~\ref{thm:LowerboundsVCsummary} proves lower bounds for \textsc{Odd Cycle Transversal~[VC]}, \textsc{Even Cycle Transversal~[VC]}, \textsc{Feedback Vertex Set~[VC]}, and \textsc{Cograph Deletion~[VC]}.
Examples for which Corollary~\ref{thm:LowerboundsVCsummary} does not give a lower bound include \textsc{Cluster Vertex Deletion~[VC]} (indeed, then a kernel is known~\cite{SaketVCNr}), or more generally, \textsc{$H$-free Deletion~[VC]} when $H$ is a star.

%% file: ms_conclusion.tex
\section{Conclusion}

We have seen different streaming algorithms and lower bounds for \textsc{$\Pi$-free Deletion} and its more specific forms, making use of the minimum vertex cover as a parameter. We have seen the potency of the AL streaming model in combination with the vertex cover, where in other streaming models lower bounds arise. It is interesting that for very local structures like a $P_3$, this combination works effortlessly, giving a very efficient memory-optimal algorithm. For more general structures troubles arise, but nonetheless, we can solve the more general problems with a many-pass, low-memory approach. Alternatively, the adaptations of kernels gives rise to a few-pass, high-memory algorithm, which provides a possible trade-off when choosing an algorithm.

We also propose the following open problems.
Can lower bounds be found expressing a pass/memory trade-off in the vertex cover size for the \problemPfDV{} problem?
Or alternatively, can we find an upper bound for \problemPfDV{} using $\O(K \log n)$ bits of memory but only a polynomial in $K$ number of passes?
Essentially, here we ask whether or not our algorithm is reasonably tight, or can be improved to only use a polynomial number of passes in $K$. A lower bound expressing a trade-off in terms of the vertex cover size is a standalone interesting question, as most lower bound statements about streaming algorithms express a trade-off in terms of $n$.

We also ask about the unparameterized streaming complexity of \textsc{Cluster Vertex Deletion} in the AL model. While lower bounds for most other \textsc{$\Pi$-free Deletion} problems in the AL model follow from our work (Theorem~\ref{thm:PiFreeDeletionALLowerBound}) and earlier work of Bishnu et al.~\cite{SaketVCNr}, this appears an intriguing open case.

Finally, we ask if there is a $2^{o(K \log K)}$ lower bound for \problemPfDV{} when $\Pi$ is characterized by few adjacencies?